\documentclass[journal]{IEEEtran}

\ifCLASSINFOpdf
\else
\fi

\usepackage{amsmath}
\usepackage{amsfonts}
\usepackage{mathtools}
\usepackage{amssymb} 
\usepackage{graphicx}
\usepackage{caption}
\usepackage{bbm}
\usepackage{gensymb}
\DeclarePairedDelimiter{\ceil}{\lceil}{\rceil}
\usepackage{flushend}
\usepackage{algorithmic}
\usepackage{algorithm}
\begin{document}
%

\title{Learned-SBL: A Deep Learning Architecture for Sparse Signal Recovery}
%
%
%

\author{Rubin Jose Peter,
	Chandra R. Murthy,~\IEEEmembership{Senior~Member,~IEEE,}
}

\markboth{Journal of XXXX ,~Vol.XXX, No.XX, XXX~XXX}%
{Shell \MakeLowercase{\textit{et al.}}: Bare Demo of IEEEtran.cls for IEEE Journals}
%



\maketitle

\begin{abstract}
In this paper, we present a computationally efficient sparse signal recovery scheme using Deep Neural Networks (DNN). The architecture of the introduced neural network is inspired from sparse Bayesian learning (SBL) and named as Learned-SBL (L-SBL). We design a common architecture to recover sparse as well as block sparse vectors from single measurement vector (SMV) or multiple measurement vectors (MMV) depending on the nature of the training data. In the MMV model, the L-SBL network can be trained to learn any underlying sparsity pattern among the vectors including  joint sparsity, block sparsity, etc.  In particular, for block sparse recovery, learned-SBL does not require any prior knowledge  of block boundaries. In each layer of the L-SBL, an estimate of the signal covariance  matrix is obtained as the output of a neural network. Then a maximum a posteriori (MAP) estimator of the unknown sparse vector is implemented  with non-trainable parameters.   In many applications, the measurement matrix may be time-varying. The existing DNN based sparse signal recovery schemes demand the retraining of the neural network using  current measurement matrix.  The  architecture of  L-SBL allows it to accept the  measurement matrix as an input to the network, and thereby avoids the need for retraining. We also evaluate the performance of Learned-SBL in the detection of an extended target using a multiple-input multiple-output  (MIMO) radar.  Simulation results illustrate that  the proposed approach offers superior sparse recovery performance compared to the state-of-the-art methods.
\end{abstract}

\begin{IEEEkeywords}
Sparse signal recovery,  Deep Neural Networks (DNN), sparse Bayesian learning (SBL)
\end{IEEEkeywords}

%
\IEEEpeerreviewmaketitle

\section{Introduction}
\IEEEPARstart{O}{ne}  of the main drawbacks of present-day sparse signal recovery algorithms is their iterative nature and computational complexity, especially in high-dimensional settings. This limits their applicability in practical scenarios. A different challenge arises in applications where it is expensive or time-consuming to acquire measurements. Recent advances in deep neural networks (DNNs) provide the tantalizing possibility of designing fixed-complexity algorithms by learning to invert the NP hard problem of finding the sparsest solution to an under-determined set of linear equations, and this paper presents a computationally efficient deep learning architecture named as Learned-SBL (L-SBL)  to accomplish this objective. Moreover, the introduced DNN architecture can recover sparse, block sparse, joint sparse or other structured sparse models   from the single or multiple measurement vectors.   

 In recent literature, DNN based sparse signal recovery has been explored, typically, by unfolding the iterations involved in existing sparse signal recovery algorithms. 
 For example, the learned coordinate descent (LCoD) and learned iterative shrinkage-thresholding algorithm (LISTA) approximate CoD and ISTA  using a DNN with a 
 specific architecture and fixed depth \cite{gregor_ICML_2010}.
 The concept of deep unfolding was introduced to obtain the advantages of both model based methods and DNNs \cite{hershey_arXiv_2014}.
 A deep learning architecture, which outperforms ISTA  by unfolding the approximate message passing (AMP) algorithm and vector AMP was introduced in  \cite{Schniter_TSP_2017}.
 Using the concept of unfolding, denoising-based approximate message passing (D-AMP) algorithm was approximated as Learned D-AMP \cite{mousavi_ANIPS_2017}. The capability of a DNN to outperform  even the algorithms in sparse signal recovery on which it is based was demonstrated theoretically and empirically in \cite{wipf_ANIPS_2016}. The connection between sparse Bayesian learning (SBL) and long short-term memory (LSTM) networks was explored in \cite{wipf_ANIPS_2017}. Approximating an iterative algorithm using a deep neural network to reduce the computational complexity was demonstrated in \cite{Sidiropoulos_SPAWC_2017}. Most of these DNN based sparse signal recovery schemes, directly or indirectly, are inspired by an existing algorithm in sparse signal processing.
 
 There are also examples of DNN architectures that are not based on sparse signal recovery algorithms. A deep learning framework based on stacked denoising autoencoder (SDA)  was introduced in \cite{mousavi_AAC_2015}, which supports both linear and mildly nonlinear measurements. Majority voting neural networks  in the binary compressed sensing  problem was proposed in  \cite{ito_arXiv_2016}, where the output of independently trained feed forward neural networks are combined to obtain an estimate of a binary sparse vector. A computationally efficient approach to learn the sparse representation and recover the unknown signal vector using  a deep convolution network was proposed in \cite{mousavi_ICASSP_2017}. An approach of sparse signal recovery using GANs was proposed in \cite{Bora_ICML_2017}, where an additional optimization problem is solved to find a suitable vector in the latent space to generate the desired sparse vector corresponding to the observation. A cascaded DNN architecture to solve the sparse signal recovery problem was introduced in \cite{zhang_ICPR_2018}.  In \cite {limmer_TSP_2018}, a theoretical framework to design neural architectures for Bayesian compressive sensing was presented.
 
 In many real world applications, the nonzero elements in the sparse vector are clustered. For example, the detection of an extended target  using a MIMO radar can be formulated as the recovery of a  block sparse signal vector. Different algorithms were proposed in the sparse signal processing literature to recover block sparse vectors. Many existing algorithms assume some prior knowledge about the block boundaries and block sizes. Algorithms like Model-CoSaMp \cite{baraniuk_IT_2008}, Block-OMP \cite{eldar_TSP_2010},  Group Basis Pursuit \cite{van_SIAM_2008}, block-sparse Bayesian learning (BSBL) \cite{zhang_TSP_2013} etc, assume  prior knowledge of the block partitions. Such algorithms are sensitive to the mismatches in the assumed block boundaries. Algorithms like pattern-coupled sparse Bayesian learning (PC-SBL) \cite{Fang_TSP_2015}, expanded block sparse Bayesian learning (EB-SBL) \cite{zhang_TSP_2013}  do not require any prior knowledge about  block boundaries. PC-SBL showed superior performance over EB-SBL. However, PC-SBL needs  sub-optimum selection of the assumed model parameters and solution, as a closed form solution to the underlying optimization  problem is not available. An alternative to the EB-SBL and its relation with PC-SBL is presented in \cite{wang_TSP_2018}.
 
 However, these algorithms are iterative in nature and computationally expensive. A deep learning based approach to enhance the performance of block sparse signal recovery with reduced  computational complexity  is less explored in  literature. A few DNN based architectures are proposed to recover sparse vectors with multiple measurement vector (MMV) model.  A DNN based channel estimation scheme using MMV model is  presented in \cite{mohades_arXiv_2018}. An LSTM based sparse signal recovery scheme for MMV model is explored in \cite{palangi_TSP_2016}.
 
 In the field of wireless communication, the measurement matrix connecting between the sparse vector and observation vector may depend on the channel between the transmitter and receiver. For example, in \cite{Du_TWC_2018},  multiuser detection  in wireless communication is formulated as a block sparse signal recovery problem and the measurement matrix depends on the channel state information. Most of the DNN based sparse signal recovery schemes existing in literature assume that the observation vector is generated from a fixed measurement matrix. If the measurement matrix changes, the DNN should be trained again with new training data. This training procedure is computationally expensive  and can not be done in real-time. 
 Thus, the main drawbacks of the existing DNN based sparse signal recovery schemes are,
 \begin{itemize}
 	\item Deep learning based block sparse signal recovery schemes without any prior knowledge of block partition using SMV or MMV models are well explored.
 	\item Existing deep learning architectures  are not suitable in the applications where  measurement matrix changes for each measurement acquired.
 \end{itemize}
 
In this paper, we present a computationally efficient DNN architecture to recover the sparse,  block sparse  as well as jointly sparse vectors. Our DNN architecture is inspired from the sparse Bayesian learning algorithm and we name the resulting DNN as Learned-SBL (L-SBL). Each layer of  L-SBL is similar to an iteration of SBL. The outputs of an L-SBL layer are the estimate of the sparse vector and the diagonal elements of the error covariance matrix. An L-SBL layer comprises two stages. In the first stage, the signal covariance matrix is estimated using the diagonal elements of the error covariance matrix and the estimate of the sparse vector at the  output of the previous layer. In the second stage, a MAP estimate of the sparse vector and error covariance matrix are implemented using non-trainable parameters.   In L-SBL, any dependency of the measurement vectors on the underlying structure within or among the sparse vectors is captured in the MAP estimation stage by the neural network  used in the estimation of  signal covariance matrix. Since the measurement matrix is  used  only in the MAP estimation stage without any trainable parameters, the L-SBL can be trained  with a randomly drawn measurement matrix.  Therefore, L-SBL can be effectively used in many scenarios where measurement matrix can be arbitrary and different across the multiple measurements. Further, L-SBL  can utilize single or multiple measurement vectors  during the recovery of sparse vectors. For example, if we train the neural network with single measurement vector,  L-SBL behaves as a sparse recovery algorithm similar to basic SBL. If the training data contains block sparse vectors, the L-SBL becomes a block sparse recovery algorithm.  That is, L-SBL can learn any underlying structure in the training dataset. Further, L-SBL provides a computationally efficient recovery scheme compared to the corresponding iterative algorithms like SBL, PC-SBL, M-SBL etc.   
Our main contributions as follows:

\begin{itemize}

	\item We design a deep learning architecture  named as Learned-SBL for different sparse signal recovery applications. Based on the nature of the training data,  L-SBL  can recover sparse, jointly sparse or block sparse vectors from single or multiple measurements.

	\item We compare the performance of L-SBL with other algorithms and show the capability of L-SBL to avoid retraining in the scenarios where measurement matrix  is different across the multiple measurements or in scenarios where the specific measurement matrix is not available during the training phase of the DNN.
	 
	\item We evaluate the capability of L-SBL to utilize any existing sparsity pattern among the nonzero elements of the source vectors to enhance the performance.
	
	\item We examine the weight matrix learned by the L-SBL network in different scenarios. This provides insight into how L-SBL is able to adapt  to different underlying sparsity patterns.
	
	\item We evaluate the performance of L-SBL  in the detection of an extended target using MIMO radar.

\end{itemize}
  
The rest of the paper is organized as follows. The problem formulation and an overview of SBL, M-SBL and PC-SBL algorithms are presented in section \ref{Problem_Form}. In section \ref{L-SBL Architecture}, we present our  Learned-SBL architecture. We introduce two architectures for the L-SBL layer and we  compare the computational complexity of an L-SBL layer with an iteration of the SBL algorithm. We also describe the training algorithm for L-SBL.    
 Numerical simulation results illustrating the performance of L-SBL in recovering sparse as well as block sparse vectors are presented in section \ref{Numerical_Analysis}. In section \ref{Extended_Detection}, the extended target detection using MIMO radar is formulated as a block sparse signal recovery problem and the performance of L-SBL in target detection is evaluated. We offer some concluding remarks in section \ref{Conclusion}.  
 
 Throughout the paper, bold symbols in small and capital letters
 are used for vectors and matrices, respectively. $x_i$ denotes the $i^{\text{th}}$ element of the vector $\mathbf{x}$. $[\mathbf{A}]_{i,j}$ represents the ${(i,j)}^{\text{th}}$ element of the matrix $\mathbf{A}$.  For the matrix $\mathbf{A}$, $\mathbf{A}_{i.}$ indicates the $i^{\text{th}}$ row of $\mathbf{A}$ and $\mathbf{A}_{.i}$ indicates the $i^{\text{th}}$ column of $\mathbf{A}$. $\mathrm{Tr}()$ indicates the trace of a matrix.   The $\ell_2$ and $\ell_0$ norm of the vector $\mathbf{x}$ are denoted by $\lvert\lvert \mathbf{x} \rvert\rvert_2$ and $\lvert\lvert \mathbf{x} \rvert\rvert_0$, respectively. For a matrix $\mathbf{A}$, $\mathbf{A}^T$ and $\mathbf{A}^{-1}$ denote the transpose and the inverse of the matrix, respectively. For a vector $\mathbf{x}$,  $diag{(\mathbf{x})}$ denotes a diagonal matrix with the elements of vector $\mathbf{x}$ as the diagonal entries. For a matrix $\mathbf{A}$, $diag{(\mathbf{A})}$  denotes the column vector containing diagonal elements of  $\mathbf{A}$.  $\mathcal{N}()$ denotes the multivariate Gaussian distribution and $\Gamma ()$ denotes the Gamma function.

\section{Problem Formulation}\label{Problem_Form}
We consider the problem of sparse signal recovery from $L$ measurement vectors $\{\mathbf{y}_l\}_{l=1}^L$, where  $\mathbf{y}_l \in \mathbb{R}^{ M \times 1} $ is related to the sparse vector $\mathbf{x}_l \in  \mathbb{R}^{ N \times 1}  $ by  the expression
 \begin{equation}
 \mathbf{y}_l = \mathbf{A} \mathbf{x}_l +\mathbf{n}_l.
 \label{eq:MeasModel}
  \end{equation}
  The above expression can be rearranged as
  \begin{equation}
  \mathbf{Y} = \mathbf{A} \mathbf{X} +\mathbf{N},
  \label{eq:MeasModel_MMV}
  \end{equation}
where $\mathbf{A} \in \mathbb{R}^{ M \times N}$ denotes the known measurement matrix with  $ M \ll N$, $\mathbf{Y} = [\mathbf{y}_1, \mathbf{y}_2 , \ldots , \mathbf{y}_L] \in \mathbb{R}^{ M \times L}$ denotes the matrix with multiple measurement  vectors and $\mathbf{X}$ represents the matrix with the sparse vectors $\{\mathbf{x}_l\}_{l=1}^L$ as its columns.  If $\mathbf{x}_l$ is a usual sparse vector, then  $\mathbf{x}_l$ contains a few arbitrarily located nonzero elements. In the MMV model, we say that  $\{\mathbf{x}_l\}_{i=l}^L$  are jointly sparse,  if the  vectors  $\{\mathbf{x}_l\}_{l=1}^L$   share a common support. In the block-sparse case, the nonzero elements  of $\mathbf{x}_l$  occur in a small number  of clusters. We assume that the noise matrix $\mathbf{N}$ contains independent and identically distributed  Gaussian random variables with zero mean and  known variance $\sigma^2$, denoted as 
$\mathbf{N}_{i,j} \sim \mathcal{N}(0,\sigma^2),\ \  i = 1,2, \dots, M, \ \ j= 1,2, \dots, L$.
In Bayesian learning, one seeks the maximum a posteriori (MAP) estimate of $\mathbf{x}_{l}$, given by:

 \begin{equation}
 \hat{\mathbf{x}}_l = \underset{\mathbf{x}_l}{\arg\max}\  p(\mathbf{x}_l \vert \mathbf{y}_l; \sigma^2)
 \label{eq:MAP_Exp}
 \end{equation}
In Bayesian learning, the prior distribution on $\mathbf{x}_l$ as modeled as zero mean Gaussian with covariance matrix $\mathbf{R}_\mathbf{x}$. Then, the posterior distribution of $\mathbf{x}_l$ is also Gaussian and the MAP estimate of $\mathbf{x}_l$ is the posterior mean:

 \begin{equation}
	\hat{\mathbf{x}}_l= \mathbf{R}_\mathbf{x}\mathbf{A}^T(\mathbf{A}\mathbf{R}_\mathbf{x}\mathbf{A}^T+\sigma^2\mathbf{I})^{-1}\mathbf{y}_l
 \label{eq:MAP_Est}
 \end{equation}
 Considering the  $L$ measurements together, the MAP estimate $ \hat{\mathbf{X}}$ and the error covariance matrix $ \mathbf{\Phi}$ are given by
  
 \begin{equation}
 \begin{split}
 \hat{\mathbf{X}}&= \mathbf{R}_\mathbf{x}\mathbf{A}^T(\mathbf{A}\mathbf{R}_\mathbf{x}\mathbf{A}^T+\sigma^2\mathbf{I})^{-1}\mathbf{Y}\\
 \mathbf{\Phi} &=  \mathbb{E} \{(\mathbf{x}-\hat{\mathbf{x}})(\mathbf{x}-\hat{\mathbf{x}})^T\}\\
 &= \mathbf{R}_\mathbf{x} - \mathbf{R}_\mathbf{x}\mathbf{A}^T(\mathbf{A}\mathbf{R}_\mathbf{x}\mathbf{A}^T+\sigma^2\mathbf{I})^{-1}\mathbf{A}\mathbf{R}_\mathbf{x}\\
 \label{eq:MAP_Est_MMV}
 \end{split}
 \end{equation}
 We use \eqref{eq:MAP_Est_MMV} to get the MAP estimate of the sparse vector in each L-SBL layer.
 
\subsection{Sparse Bayesian Learning (SBL)}
Sparse Bayesian learning \cite{tipping_MLR_2001,wipf_TSP_2004} is a well known algorithm to recover a sparse vector from  under-determined set of measurements. The SBL algorithm was originally proposed to recover the sparse vector $\mathbf{x}_l$ from single measurement vector $\mathbf{y}_l$. In SBL, the sparse vector $\mathbf{x}_l$ is modeled as being Gaussian distributed with a diagonal covariance matrix
 \begin{equation}
\mathbf{x}_l  \sim \mathcal{N}(\mathbf{0},\mathbf{R}_\mathbf{x}),  
 \label{eq:SBL1}
 \end{equation}
  \begin{equation}
   \mathbf{R}_\mathbf{x} =diag([\frac{1}{\alpha_1}, \frac{1}{\alpha_2}, ....\frac{1}{\alpha_N}]),
  \label{eq:SigCorr}
  \end{equation}
  where $\alpha_i$ denotes the inverse of the variance of $i^{\text{th}}$ element of the sparse vector $\mathbf{x}_l$.  Also $\alpha_i$  is assumed to  be a  Gamma distributed random variable with parameters $a$ and $b$:
  \begin{equation}
  p(\boldsymbol{\alpha})=\prod_{i=1}^{N}Gamma(\alpha_i\vert a,b)=\prod_{i=1}^{N}\Gamma(a)^{-1}b^a\alpha_i^ae^{-b\alpha_i}.
    \label{eq:GammaModel}
  \end{equation} 
   It can be shown that, the prior distribution of $\mathbf{x}$ with respect to the parameters $a$ and $b$ is a  students-t distribution, which is known to be a sparsity promoting prior distribution.
   Specifically, for small values $a$ and $b$, the students-t distribution has sharp peak at zero, which favors sparsity. To find an estimate $\alpha_i$, a lower bound of the posterior density,     $p(\boldsymbol{\alpha} \vert \mathbf{y}_l)$  is maximized using the Expectation Maximization (EM) algorithm. This leads to an iterative recipe, where the  update of $\alpha_i$ at iteration $t$, denoted by $\alpha_i^t$ is given by
        \begin{equation}
       \alpha_i^t = \frac{1}{(x_i^{t-1})^2+ [\mathbf{\Phi}^{t-1}]_{i,i}},
   \label{eq:SBL_Update}     
        \end{equation}
  where $x_i^{t-1}$ denotes the estimate of the $i^{\text{th}}$ element of the sparse vector and $[\mathbf{\Phi}^{t-1}]_{i,i}$ is the $i^{\text{th}}$  diagonal entry of the estimated  error covariance matrix $\mathbf{\Phi}$,  in the ${t-1}^{\text{th}}$ iteration.
  
\subsection{Sparse Bayesian Learning using Multiple Measurement Vectors (M-SBL)}  
In \cite{wipf_TSP_2007}, the basic SBL algorithm is extended to handle multiple measurement vectors, resulting in the M-SBL algorithm. M-SBL reduces the failure rate and mean square error by utilizing the joint sparsity across the multiple sparse vectors. In M-SBL, each row of the matrix $\mathbf{X}$ is assumed to be distributed as a Gaussian random vector,
 \begin{equation}
 \mathbf{X}_{i.}  \sim \mathcal{N}(\mathbf{0},\alpha_i^{-1}\mathbf{I}), 
 \label{eq:SBL_MMV_prior}
 \end{equation}
  where the hyperparameters, $\{\alpha_i\}_{i=1}^N$  are   Gamma distributed similar to \eqref{eq:GammaModel}. The hyperparameters $\{\alpha_i\}_{i=1}^N$ are estimated by maximizing the posterior density  $p(\boldsymbol{\alpha} \vert \mathbf{Y})$.
 Similar  to the SBL algorithm, the update of $\alpha_i^{t}$ is obtained by maximizing a lower bound on $\log(p(\boldsymbol{\alpha} \vert \mathbf{Y}))$ using the EM algorithm, which leads to the iterative update equation given by
         \begin{equation}
         \alpha_i^t = \frac{1}{\frac{1}{L} \Vert \mathbf{X}_{i.}^{t-1} \Vert_2^2+ [\mathbf{\Phi}^{t-1}]_{i,i}}.
         \label{eq:M-SBL_Update}     
         \end{equation}
\subsection{Pattern-Coupled Sparse Bayesian Learning (PC-SBL)}
Pattern coupled sparse Bayesian learning \cite{Fang_TSP_2015} extends SBL algorithm to recover block sparse vectors when the block boundaries are unknown.  In PC-SBL, since the nonzero elements occurs as clusters, a coupling model is assumed  between the adjacent elements of the  vector. Mathematically, the diagonal elements of the  signal covariance matrix in  \eqref{eq:SigCorr} is modeled as
  \begin{equation}
  [\mathbf{R}_\mathbf{x}]_{i,i} = (\alpha_i+\beta\alpha_{i-1}+\beta\alpha_{i+1})^{-1},
  \label{eq:CouplingModel}
  \end{equation}  
where $\beta$ is the non negative coupling parameter, and $\alpha_0$ and  $\alpha_{N+1}$ are assumed to be zero. In PC-SBL, $\alpha_i$  is assumed to be a Gamma distributed  random variable with parameters $a$ and $b$, similar to  \eqref{eq:GammaModel}. The entanglement of  $\{\alpha_i\}_{i=1}^N$ through the coupling parameter $\beta$ precludes a closed form solution in the M-step of the EM algorithm. However, we can find the feasible set for the solution $\alpha_i^t$ as
  \begin{equation}
    \alpha_i^t  \in \Big[ \frac{a}{.5\omega_i^{t-1}+b},\frac{a+1.5}{.5\omega_i^{t-1}+b} \Big],
    \end{equation}
where $\omega_i^{t-1}$ is given by
\begin{equation}
 	\begin{split}
	\omega_i^{t-1} =  ({(x_i^{t-1})}^2 + [\mathbf{\Phi}^{t-1}]_{i,i})+ \beta({(x_{i+1}^{t-1})}^2 + [\mathbf{\Phi}^{t-1}]_{i+1,i+1})+\\ \beta({(x_{i-1}^{t-1})}^2  + [\mathbf{\Phi}^{t-1}]_{i-1,i-1}). 
	 	\end{split} 
 \end{equation}  
One major drawback in the PC-SBL algorithm is the  sub-optimum selection of the update equation for $\alpha_i^t$ as the lower bound of the feasible set.:
  \begin{equation}
    \alpha_i = \frac{a}{.5\omega_i^{t-1}+b}.
    \label{eq:PC_SBL}
  \end{equation}
Due to this, the convergence of the EM algorithm is no longer guaranteed. Also,  no theoretical guarantees on the performance is available, even though the algorithm  empirically offers excellent recovery performance. 
\subsection{Discussion}
Many algorithms for block sparse signal recovery require prior knowledge about block boundaries. The PC-SBL algorithm does not require any prior knowledge. However, PC-SBL assumes a coupling model in \eqref{eq:CouplingModel}, which may be sub-optimum in many practical applications. In this coupling model, $\beta=0$ leads to the original SBL algorithm and any nonzero value  of  $\beta$ leads to the adjacent elements being coupled. The optimum choice of the coupling parameter $\beta$  depends on the  nature of the block sparse vectors.   We can also consider a coupling model other than in \eqref{eq:CouplingModel}, for example
   \begin{equation}
   [\mathbf{R}_\mathbf{x}]_{i,i} = (\alpha_i+\beta_1\alpha_{i-1}+\beta_2\alpha_{i-2}+\beta_1\alpha_{i+1}+\beta_2\alpha_{i+2})^{-1}.
   \label{eq:CouplingModel2}
   \end{equation}  
The main difficulty  in using these models is the difficulty in obtaining a closed form  solution of  the hyperparameters $\{\alpha_i\}_{i=1}^N$. In PC-SBL, a sub-optimum  solution is chosen as   $\{\alpha_i\}_{i=1}^N$ from the feasible set. Other hyperparameters like $\beta$, $a$ and $b$ are also selected heuristically.   Major drawbacks of the PC-SBL algorithm are summarized below.
\begin{itemize}
	\item Coupling parameter $\beta$ and number of terms in the coupling model are selected heuristically.
	\item Sub-optimum  selection of $\alpha_i^t$  from the feasible set.
	\item Heuristic selection of  the hyperparameters $a$ and $b$.
\end{itemize}
In such scenarios,  deep learning can potentially do a better job by optimally estimating these parameters and the  coupling model from a training data set. 

Algorithms like M-SBL assume the source vectors in multiple measurements are jointly sparse. In several practical applications, the nonzero elements among the sparse vectors may not share a common support.  For example, in the direction of arrival (DoA) estimation problem, the  simultaneous presence of fast moving targets and stationary targets can create a scenario as shown in Figure \ref{MMV_Arbit_Model}. In such cases,  existing algorithms  like M-SBL fail to utilize multiple measurements to improve the performance of sparse signal recovery.   In such multiple measurements scenarios,  a DNN  can  possibly  learn an inverse function from the training data, which incorporates  arbitrary sparsity patterns among the source vectors, to improve the signal recovery performance.

From  \eqref{eq:SBL_Update}, \eqref{eq:M-SBL_Update} and \eqref{eq:PC_SBL}, we notice that the estimate of the signal covariance matrix in  $t^{\text{th}}$ iteration can be expressed as function of the sparse vector and the diagonal elements of the error covariance matrix estimated in $t-1^{\text{th}}$ iteration.  The update of the diagonal element of signal covariance matrix  is given by
  \begin{equation}
  [\mathbf{R}_\mathbf{x}]_{i,i} = f_i(x_1^{t-1},x_2^{t-1}..x_N^{t-1}, [\mathbf{\Phi}^{t-1}]_{1,1},...[\mathbf{\Phi}^{t-1}]_{N,N})
  \label{eq:DNN_Function}
  \end{equation}
where $\{ f_i\}_{i=1}^N$ depends on the nature of the sparse signal recovery problem.      
From the training data, the L-SBL can learn the functions $\{ f_i\}_{i=1}^N$,  which connects  the previous estimate of the sparse vector to the signal covariance matrix in the current iteration. Such a DNN based approach can avoid the major drawbacks of existing approaches. In the MMV model with arbitrary source patterns, L-SBL can learn more suitable functions $\{ f_i\}_{i=1}^N$ by utilizing the patterns among source vectors.   Further, the SBL algorithm was derived with the assumption that the  probability density $p(x_i \vert \alpha_i)$ is Gaussian. Deviations from the assumed model may affect the performance of the algorithm.  In such scenarios also, the L-SBL can outperform SBL with reduced computational complexity.  With this preliminary discussion of the motivation and need for more general sparse recovery techniques, we are now ready to present the L-SBL architecture in the next section.

\section{L-SBL Architecture and Training}\label{L-SBL Architecture}
\subsection{L-SBL Architecture}
\begin{figure}[t]
 	\centering	
 	\includegraphics[width=9cm,height=6cm]{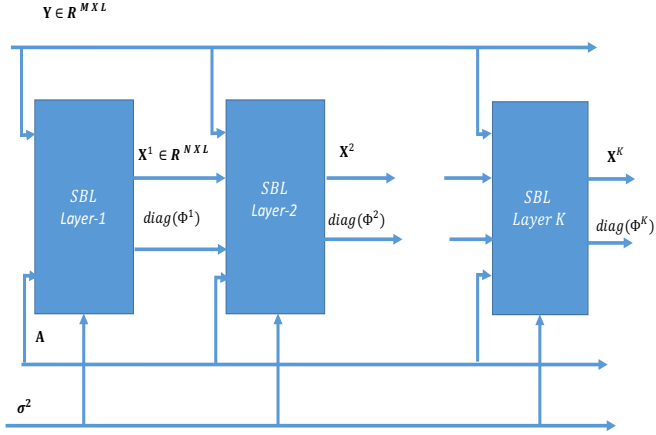} 	
 	\caption{Learned-SBL}
 	\label{L_SBL_Arch}
 \end{figure}

 The L-SBL architecture with multiple layers is shown in Figure \ref{L_SBL_Arch}. Each layer of the L-SBL is similar to an iteration of the SBL or PC-SBL algorithm. The inputs  to the  L-SBL network are  the measurement matrix $\mathbf{A}$, the  measurement vector $\mathbf{y}$ and  noise variance $\sigma^2$.  The outputs of the $t^{\text{th}}$ L-SBL layer are the estimate of the sparse vector $\mathbf{x}^t$ and the diagonal elements of the error covariance matrix $diag(\mathbf{\Phi^t})$.
A single layer of L-SBL network has two stages.  In the first stage, we have a neural network which accepts  the estimate of  sparse vector $\mathbf{x}$ and diagonal elements of the error covariance matrix  $\mathbf{\Phi}$ from the previous layer and gives the diagonal elements of the signal covariance matrix $\mathbf{R}_\mathbf{x}$ as the outputs. The designed L-SBL architecture can  learn  functions $\{ f_i\}_{i=1}^N$ similar to \eqref{eq:SBL_Update}, \eqref{eq:M-SBL_Update} or \eqref{eq:PC_SBL} depending on  the nature of the training data. The DNN can also learn a better mapping which  minimizes the mean square error between the sparse vector estimated by L-SBL and true sparse vector. 

The second stage of the  L-SBL layer gives an estimate of the sparse vector and  error covariance matrix  using   \eqref{eq:MAP_Est}. The output of the neural network in the first stage is used in the second stage for MAP estimation. The second stage of the L-SBL layer does not contain any trainable parameters. 

We consider two different neural network architectures in the design of  L-SBL layer. These two architectures are identical in single measurement vector (SMV) model. In MMV model, the second architecture vectorizes the observation matrix $\mathbf{Y}^T$ in to a vector $\mathbf{y} = vec({\mathbf{Y}^T})$ and the measurement matrix is modified as $\hat{\mathbf{A}} = \mathbf{A} \otimes \mathbb{I}_L$. We describe the two architectures below. 

\begin{figure}[t]
 	\centering
 	\includegraphics[width=9cm,height=6cm]{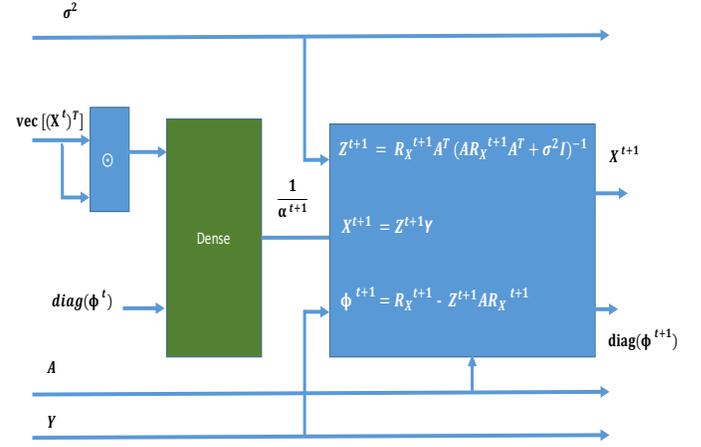} 	
 	\caption{Single Layer of \emph{L-SBL (NW-1)}}
 	\label{L_SBL_NW_1}
 \end{figure}

\emph{L-SBL (NW-1)}: In this architecture, we use a  dense  network to estimate the signal covariance matrix from the outputs of the previous layer. The number of input nodes to the dense network is $NL+N$, where $NL$ input nodes accept  $vec{(\mathbf{X}^T)^{t-1}} \odot vec{(\mathbf{X}^T)^{t-1}}$ and $N$ input nodes accept the diagonal elements of the error covariance matrix $diag{(\Phi)}^{t-1}$. We can use a single or multi layer dense network in the estimation of the hyperparameters $\{\alpha_i\}_{i=1}^N$. In our numerical studies, we consider a single layer dense network. The details of the \emph{L-SBL (NW-1)} architecture are shown in Figure \ref{L_SBL_NW_1}. In L-SBL (NW1), the number of hyperparameters $\alpha_i$ is $N$. Note that, in the second stage,  \eqref{eq:MAP_Est} is used for  MAP estimation,  which requires the inversion of a matrix of dimension $M \times M$. 

\begin{figure}[t]
 	\centering
 	\includegraphics[width=9cm,height=6cm]{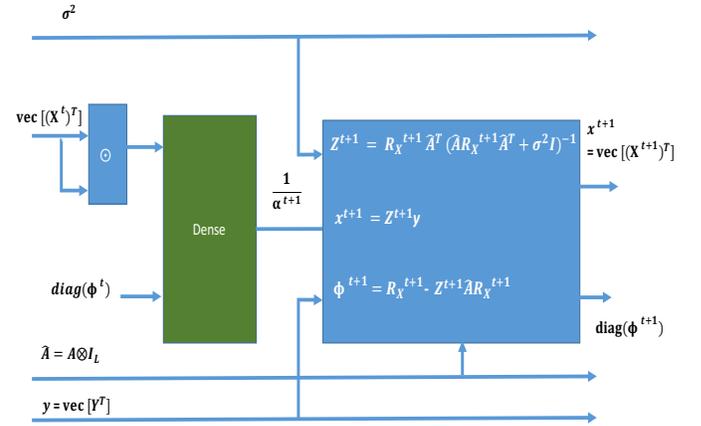} 	 	
 	\caption{Single Layer of \emph{L-SBL (NW-2)}}
 	\label{L_SBL_NW_2}
 \end{figure}

\emph{L-SBL (NW-2)}: In the second architecture, the number of hyperparameters (output nodes)  in the first stage of each layer is $NL$. The number of input nodes to the dense network is $NL+NL$, with  $vec{(\mathbf{X}^T)^{t-1}} \odot vec{(\mathbf{X}^T)^{t-1}} \in \mathbb{R}^{NL \times1}$  and $diag{(\Phi)}^{t-1} \in \mathbb{R}^{NL \times 1}$ being the inputs to the dense network. Details of the \emph{L-SBL (NW-2)} architecture are shown in Figure \ref{L_SBL_NW_2}. The measurement matrix is modified as $\hat{\mathbf{A}} = \mathbf{A} \otimes \mathbb{I}_L$, and we vectorize the observation matrix $\mathbf{Y}$. Here, the  MAP estimation stage requires the inversion of a matrix of dimension $ML \times ML$.\\

\emph{L-SBL (NW-2)} is  computationally more expensive than \emph{L-SBL (NW-1)}. The number of hyper parameters $\alpha_i$ and the dimension of the modified  measurement matrix $\hat{\mathbf{A}}$ is increased by a factor $L$. However, the  \emph{L-SBL (NW-2)} has more degrees of freedom than \emph{L-SBL (NW-1)}. Therefore, the \emph{L-SBL (NW-2)} can improve the signal recovery performance in scenarios where the nonzero elements among source vectors follow   arbitrary patterns (see  Figure \ref{MMV_Arbit_Model}).  

\subsection {Computational Complexity}
 The  MAP estimation is the common step in each layer of L-SBL network and in each iteration of the algorithms like SBL, M-SBL or  PC- SBL. In MAP estimation stage,  the inversion of an $M \times M$ matrix is required.  Matrix inversion is a computationally expensive  mathematical operation, which  requires $\mathcal{O}(M^3)$ floating point operations. SBL, M-SBL and PC-SBL algorithms have computationally simple expressions to estimate the signal covariance matrix from  the sparse vector  and the diagonal elements of the error covariance matrix  at the output of the previous iteration.  Therefore, the MAP estimation is the computationally expensive step in these algorithms.

In L-SBL, we use a neural network to estimate the signal covariance matrix. As long as the number of floating point operations in  the neural network is less than  that in the MAP estimation step,  the computational complexity of the MAP estimation stage dominates. Then, one layer of L-SBL  and one iteration of  algorithms like SBL  have  same complexity. For example, if we use a single layer dense network in the estimation of the signal covariance matrix, then each layer of  the L-SBL network needs $\mathcal{O}(N^2)$ multiplications and additions.  As long as  $N^2$ is the  less than $M^3$,  computational complexity of an  L-SBL layer and an iteration of the algorithms like SBL are of the same order. In the case of MMV model, the architecture \emph{L-SBL (NW2)}  requires a matrix inverse with dimension $ML \times ML$. Therefore, the architecture \emph{L-SBL (NW2)} is computationally more expensive than  \emph{L-SBL (NW1)}. For jointly sparse source vectors,  \emph{L-SBL (NW1)} is sufficient to reduce mean square error and failure rate. If the computational complexity in the MAP estimation stage dominates,  then  the \emph{L-SBL (NW1)}  and M-SBL have similar  complexity. In our numerical simulations, we consider a single layer dense network in each L-SBL layer. Therefore, the computational complexity of an L-SBL layer and an iteration of the SBL algorithm are comparable.

\subsection{Training of L-SBL}
Since we know the  model connecting  the unknown sparse vectors to the measurement vectors, we can train the L-SBL network using a synthetically generated  data set. Training using synthetic data set was  followed in many existing DNN based sparse signal recovery schemes \cite{gregor_ICML_2010,Schniter_TSP_2017,Sidiropoulos_SPAWC_2017}.  The algorithm used to train L-SBL  is presented in Algorithm \ref{Alg1},  which is similar to the training scheme of Learned-VAMP in \cite{Schniter_TSP_2017}. Each layer of the L-SBL network is trained one after another using the loss function given in  \eqref{eq:LossFunction}. The training of each layer has two phases. In the first phase, the trainable parameters in the previous layers are not allowed to change,  and the parameters in the current layer are updated using the training data. In the second phase, we update all trainable parameters from the first layer to the current training layer.    In Algorithm \ref{Alg1}, $\mathbf{W}_k$ denotes the set of all trainable parameters in $k^{\text{th}}$ layer. The total number of  L-SBL layers is denoted as $K$ and $R$ denotes the total number of mini-batches used in the training of an L-SBL layer. The measurement vector $\mathbf{y}_i$ is related to the sparse vector $\mathbf{x}_i$ by  
\begin{equation*}
\mathbf{y}_i = \mathbf{A}_i \mathbf{x}_i
\end{equation*}
The loss function used in the training of the L-SBL layer is the mean square error between the  true sparse vector and the current training layer output. The expression for mean square error loss function is given by

\begin{equation}
\mathcal{L} = 	\frac{1}{m}\sum_{i=1}^m\lvert\lvert{\mathbf{x}_i-\mathbf{G}(\mathbf{y}_i,  \mathbf{A}_i, \sigma^2)}\rvert\rvert_2^2, 	
\label{eq:LossFunction}
\end{equation}  
where $m$ denotes the number of training samples in a mini-batch  and $\mathbf{G}$ represents  the function learned by L-SBL. The L-SBL network is implemented  in Python using the neural network libraries Keras and tensorflow \cite{KerasText}. 

\begin{algorithm}[t]\caption{Learned-SBL  Training}\label{Alg1} \begin{algorithmic}[1] \STATE Input:, Input: Training set $\{\mathbf{y}_i,\mathbf{A}_i,\mathbf{x}_i\}_{i=1}^S$, noise variance $\sigma^2$. 	\\
		\STATE Initialize:  $\mathbf{\Theta}_0 = \{\mathbf{W}_0\}$ \\		\FOR {$k=1$ to $k=K$}\small
		
		\STATE Initialize: $\mathbf{W}_k = \mathbf{W}_{k-1} $

		\FOR{$R$ steps }
		
		\STATE Sample mini-batch of $m$ measurement vectors
		${\{\mathbf{y}_1,\mathbf{y}_2..\mathbf{y}_m\}}$ from the training set
		\STATE Fix $\mathbf{\Theta}_{k-1} = \{\mathbf{W}_t\}_{t=1}^{k-1}$ 
		\STATE Learn $\mathbf{W}_k$ by minimizing the loss function
		
		\begin{equation} \label{eq:SBL_Algo1}
		\begin{split}
		\frac{1}{m}\sum_{i=1}^m\lvert\lvert{\mathbf{x}_i-\mathbf{G}(\mathbf{y}_i,  \mathbf{A}_i, \sigma^2)}\rvert\rvert_2^2 		
		\end{split}
		\end{equation}
		
		\ENDFOR
		
		\FOR{$R$ steps }
		
		\STATE Sample mini-batch of $m$ measurement vectors
		${\{\mathbf{y}_1,\mathbf{y}_2..\mathbf{y}_m\}}$ from the training set
		\STATE Re learn $\mathbf{\Theta}_{k} = \{\mathbf{W}_t\}_{t=1}^{k}$  by minimizing the loss function

		\begin{equation} \label{eq:SBL_Algo2}
		\begin{split}
		\frac{1}{m}\sum_{i=1}^m\lvert\lvert{\mathbf{x}_i-\mathbf{G}(\mathbf{y}_i,  \mathbf{A}_i, \sigma^2)}\rvert\rvert_2^2 		
		\end{split}
		\end{equation}
		
		\ENDFOR 		
		
		\ENDFOR \end{algorithmic} \end{algorithm}

In this section, we discussed the architecture of the L-SBL network. Each layer of L-SBL comprises a hyperparameter estimation stage  using a neural network and MAP estimation stage. The MAP estimation stage does not contain any trainable parameters. The presented architectures, \emph{L-SBL (NW1)} and \emph{L-SBL (NW2)} differ only in the MMV model and \emph{L-SBL (NW2)}  has more degrees of freedom compared to \emph{L-SBL (NW1)}.  We also described the training procedure of L-SBL network using synthetically generated training data set. In the next section, we evaluate the performance of the proposed L-SBL network in different scenarios.
\section{Numerical Analysis}\label{Numerical_Analysis}
In this section, we numerically evaluate the performance of L-SBL and compare it against other state-of-the-art algorithms in the literature. We  compare the performance of  L-SBL in the recovery of usual (unstructured) sparse vectors with  existing algorithms in \ref{Simulation_Sparse}. Later,  we explore the potential of the L-SBL network to recover  block sparse vectors in \ref{Simulation_Block_Sparse}.  In \ref{Simulation_Arbitrary_Matrix},  we evaluate the capability of L-SBL to avoid retraining in the scenarios where  measurement matrix changes.   We also demonstrate that L-SBL can exploit multiple measurements with jointly sparse source vectors to improve the recovery performance and corresponding simulation results are presented in \ref{Simulation_MMV_Model}. Finally, we  simulate the source vectors with  arbitrary patterns  among nonzero elements as shown in Figure \ref{MMV_Arbit_Model} and the signal recovery performance is illustrated in  \ref{Simulation_Arbitrary_Pattern}. Numerical evaluation shows that L-SBL can utilize these source patterns in the training data set to learn a better inverse function and outperforms  the existing algorithms like SBL, M-SBL, etc.
In \ref{Weight_Matrix_Analysis},  we analyze the weight matrices learned by the L-SBL network in different scenarios like the recovery of  sparse vectors, block sparse vectors, etc. In the simulation studies, we consider a single layer dense network in each L-SBL layer.  Therefore, the computational complexity of an L-SBL layer and an iteration of the SBL algorithm are in the same order.

The relative mean square error (RMSE) and failure rate are the two metrics considered to compare the performance of different algorithms. Let $\hat{\mathbf{x}}$ be the signal recovered by a sparse recovery algorithm. The relative mean square error is given by
\vspace{-0.1cm}  
\begin{equation}
\begin{split}
\textbf{RMSE} = \frac{1}{P}\sum_{p=1}^{P}{\frac{ \lvert\lvert \mathbf{x}_p - \hat{\mathbf{x}}_p \rvert\rvert_2 ^2 }{ \lvert\lvert \mathbf{x}_p  \rvert\rvert_2 ^2}}.
\end{split}
\end{equation}	
The probability of success in the support recovery of the $p^{\text{th}}$ measurement vector is computed as
\vspace{-0.1cm}
\begin{equation}
\begin{split}
P_{p} = \frac{  \lvert \hat{S}_p \cap S_{p} \rvert }{ \lvert\lvert \mathbf{x}_p \rvert\rvert _0}, 
\end{split}
\end{equation}
where $S_p$ and $\hat{S}_p$ denote the support of $\hat{\mathbf{x}}_p$ and $\mathbf{x}_p$, respectively, and $|\cdot|$ represents the cardinality of the set. The support recovery failure rate is computed as 
\vspace{-0.1cm}	
\begin{equation}
\begin{split} 
\textbf{F}_{r} = \frac{1}{P} \sum_{k=1}^{P} \mathbbm{1}_{\{ P_{p} \neq 1 \}}
\end{split}
\end{equation}	
where $\mathbbm{1}_{\{ \}}$ denotes the indicator function.

\subsection{Sparse Signal Recovery}\label{Simulation_Sparse}
\begin{figure}[t]
	\includegraphics[width=9cm,height=6cm]{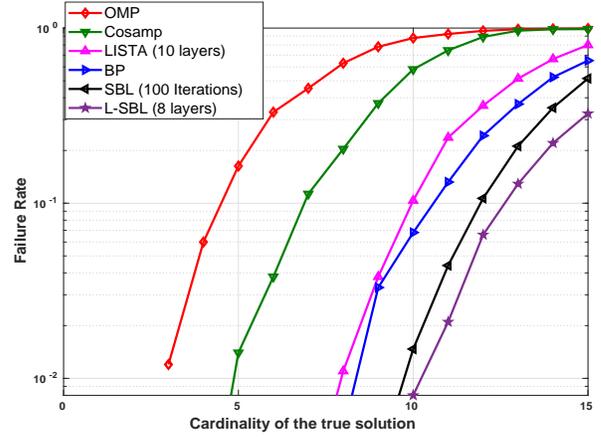} 	
	\caption{Comparison of  Failure Rate (Sparse Signal Recovery) }
	\label{Sparse_FR_1}
\end{figure}

\begin{figure}[t]
	\includegraphics[width=9cm,height=6cm]{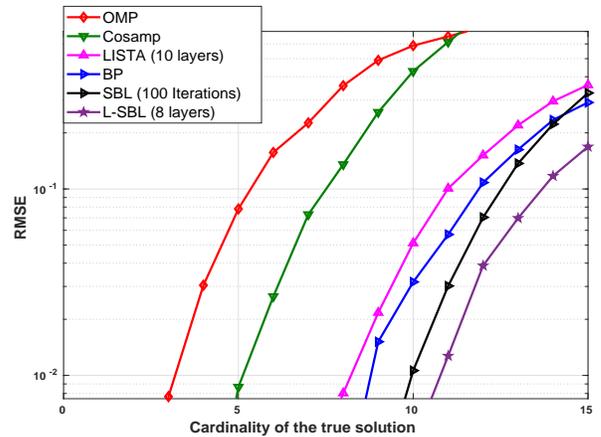} 	
	
	\caption{Comparison of RMSE (Sparse Signal Recovery) }
	\label{Sparse_MSE_1}
\end{figure}

In the first experiment, we demonstrate the performance of L-SBL network in the recovery of sparse vectors from an under-determined set of measurements. We consider a  measurement matrix $\mathbf{A}$ with dimensions $M=30$ and $N=50$. The elements of the measurement matrix $\mathbf{A}$ is drawn from Gaussian distribution with zero mean and unit variance. The maximum number of nonzero elements in the sparse vector $\mathbf{x}$ is restricted to $15$. In the testing as well as training data, the number of nonzero elements are drawn uniformly between $0$ and $15$, $\lVert \mathbf{x}  \rVert_0 \leq 15$. Amplitude of the nonzero elements is chosen  from $[.75,1] \cup [ -.75 , -1] $ with uniform probability. We consider Orthogonal Matching Pursuit (OMP), Basis Pursuit (BP), CoSamp, LISTA, sparse Bayesian learning (SBL) and L-SBL with eight layers in the comparison.   Training of the L-SBL algorithm is carried out using synthetically generated data set according to the Algorithm \ref{Alg1}. The DNN is trained using $10^6$ measurement vectors generated according to $\mathbf{y}_i = \mathbf{A}\mathbf{x}_i$. After completing the training of the DNN, algorithms are compared using a testing data set. we consider a set of $P=2000$ measurement vectors during the testing phase to evaluate the performance at each value of the sparsity level. The level is varied between $1$ and $15$.   The comparison of the failure rate and relative mean square error of different algorithms is shown in Figures \ref{Sparse_FR_1} and \ref{Sparse_MSE_1}. The plots show that L-SBL and SBL outperform the other algorithms. The relative mean square error and failure rate of the L-SBL with eight layer is less than the $100$ iterations of the SBL algorithm, indicating the computational advantage of using the DNN based approach. 
\subsection {Block Sparse Signal Recovery}\label{Simulation_Block_Sparse}
To evaluate block sparse signal recovery performance of L-SBL, we consider the same simulation parameters as in \cite{Fang_TSP_2015}. The measurement matrix $A$ has dimensions  $M=40$ and $N=100$. Let $J$ be the number of blocks in the sparse vector. The procedure to determine the block sizes $\{B_j\}_{j=1}^J$  and block boundaries is described below. 

\begin{figure}[t]
  	\includegraphics[width=9cm,height=6cm]{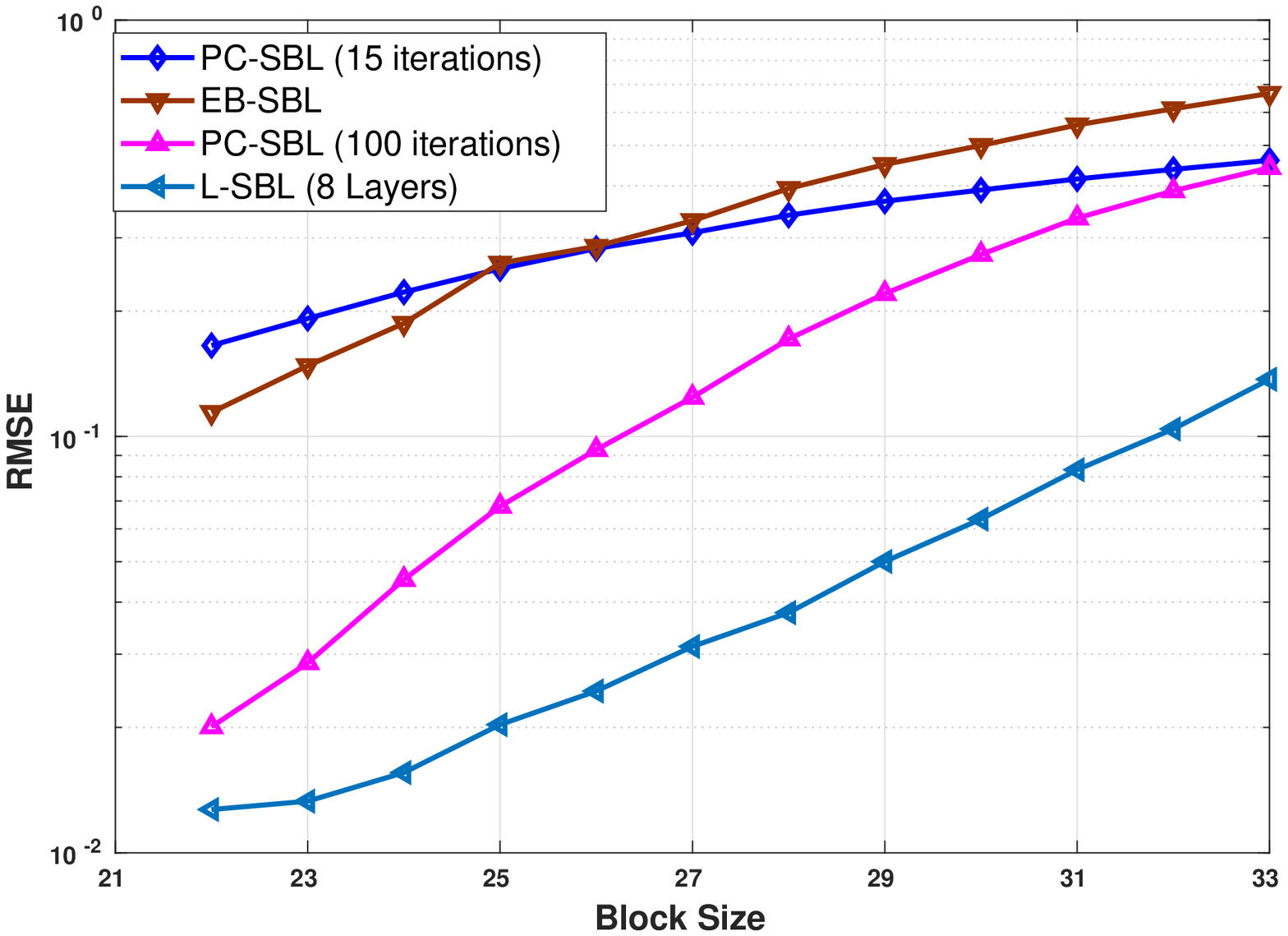} 	
  	
  	\caption{Comparison of  RMSE (Block Sparse, $J=3$) }
  	\label{Block_MSE_1}
  \end{figure}

\begin{figure}[t]
 	\includegraphics[width=9cm,height=6cm]{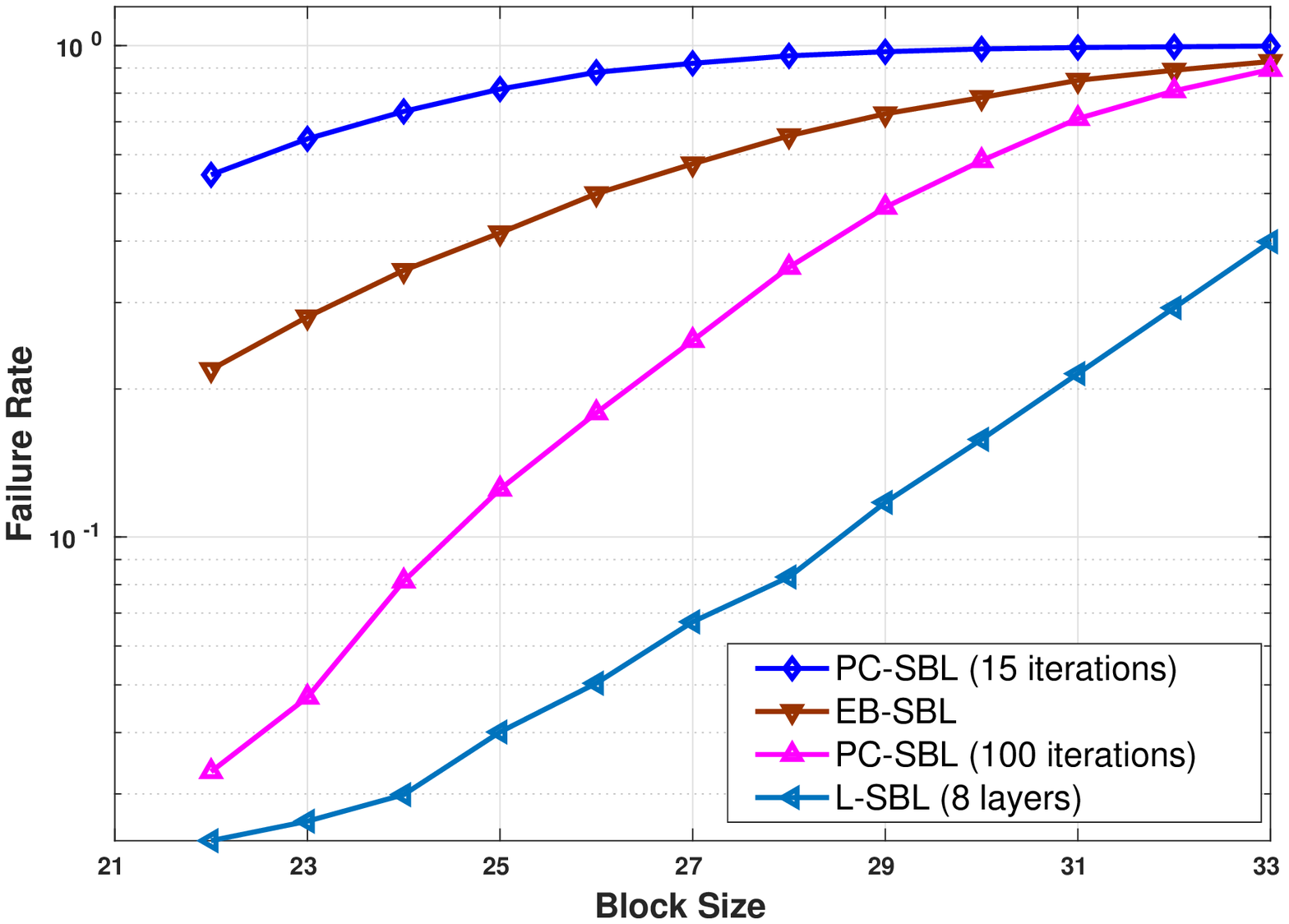} 	
 	\caption{Comparison of  Failure Rate (Block Sparse, $J=3$)}
 	\label{Block_FR_1}
 \end{figure}

Let $K$ be the number of non-zero elements in a sparse vector.  We generate $J$ positive random numbers, $\{r_j\}_{j=1}^{J}$  such that  $\sum_{i=1}^J r_j = 1$. The block size $B_j$ of the $j^{\text{th}}$ block  from $j=1$ to $j=J-1$  is chosen as $\ceil{Kr_j}$ and the block size  $B_J$  of the last block  is fixed as $K- \sum_{j=1}^{J-1}B_j$.  The locations of the non-zero blocks are also chosen randomly. First we consider $J$ partitions of the vector $\mathbf{x}$. The size of each partition is chosen as $\ceil{Nr_j}$.  Then, the $j^{\text{th}}$ nonzero block with  size $\ceil{Kr_j}$ is placed in $j^{\text{th}}$ partition of the vector $\mathbf{x}$ with a randomly chosen starting location.     In the  experiment, the maximum number of blocks, $J$ is fixed as $3$. The number of nonzero elements in the sparse vector is varied from  $21$ to $33$.  The amplitudes of the nonzero elements are chosen  from $[.75,1] \cup [ -.75 , -1] $ with uniform probability. We compare EB-SBL, PC-SBL with $15$ iterations, PC-SBL with $100$ iterations and L-SBL with eight layers. The computational complexity of L-SBL is less than that of PC-SBL with $15$ iterations. The relative mean square error is plotted as a function of  the cardinality of the true solution  in Figure \ref{Block_MSE_1}, and Figure \ref{Block_FR_1} shows the failure rate of different algorithms.
The plots illustrate the superior performance of the L-SBL over other algorithms like PC-SBL and EB-SBL.

 \subsection {L-SBL with Arbitrary Measurement Matrix}\label{Simulation_Arbitrary_Matrix} 
\begin{figure}[t]
  	\centering 	
  	\includegraphics[width=9cm,height=6cm]{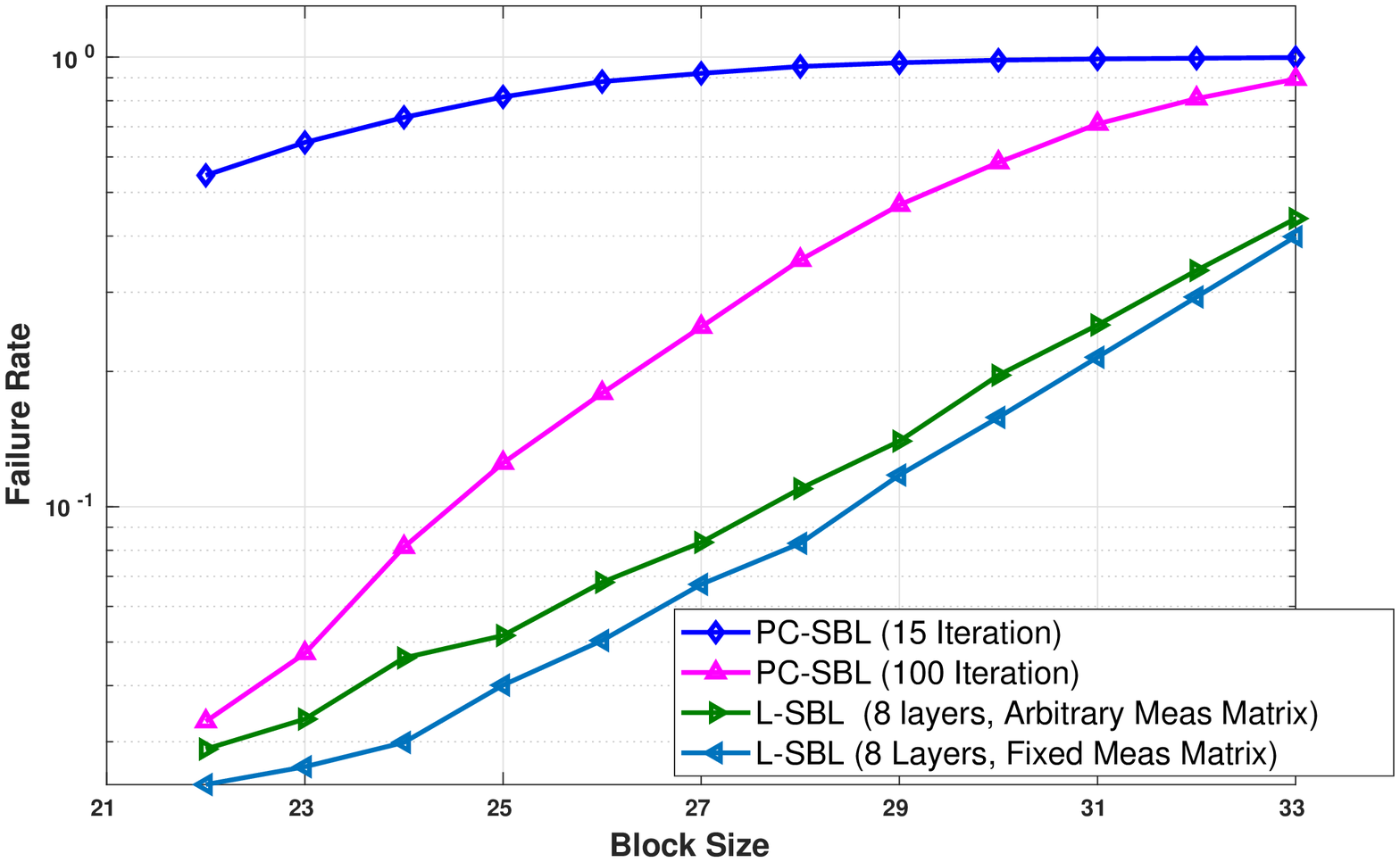} 	
  	\caption{Comparison of  Failure Rate (Block Sparse, $J=3$)}
  	\label{Block_Arbit_FR_1}
  \end{figure}

\begin{figure}[t]
 	\centering	
 	\includegraphics[width=9cm,height=6cm]{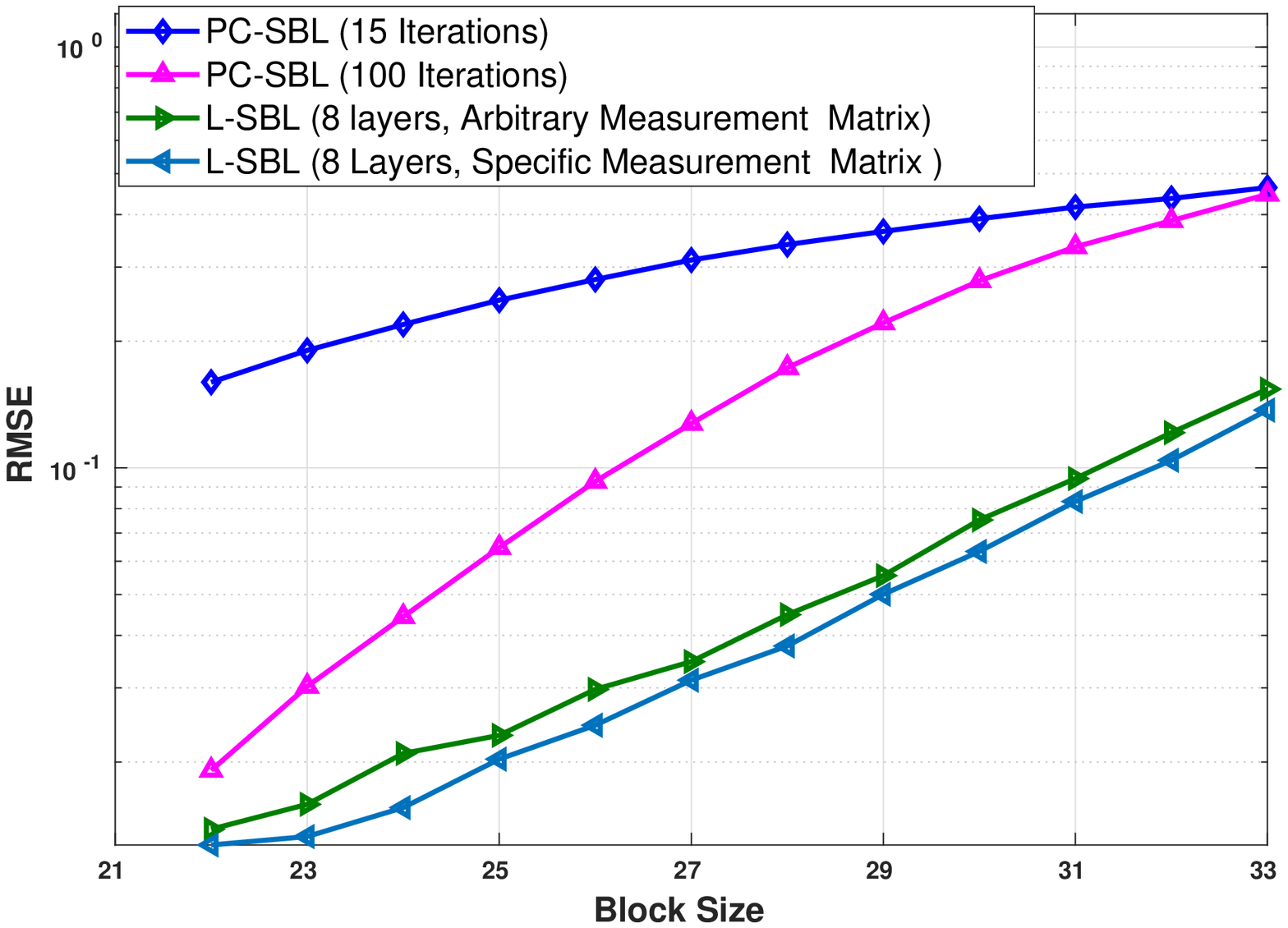} 
 	\caption{Comparison of  RMSE (Block Sparse,$J=3$) }
 	\label{Block_Arbit_MSE_1}
 \end{figure}

 The MAP estimation stage in an L-SBL layer does not contain any trainable parameters. Therefore, L-SBL can be trained using the measurement vectors from  different measurement matrices and  corresponding sparse vectors.  In each training sample, the elements of the measurement matrix $\mathbf{A}_i$ are randomly drawn from a particular  distribution  and the measurement vector $\mathbf{y}_i$ is generated as $\mathbf{A}_i\mathbf{x}_i$,  where $\mathbf{x}_i$ is a sparse or block sparse vector.  The measurement matrix and the measurement vector $\{ \mathbf{y}_i , \mathbf{A}_i \}$ are given as the inputs to  L-SBL  network. In order to  illustrate the performance of  block sparse signal recovery with arbitrary measurement matrix, randomly drawn measurement matrices $\mathbf{A}_i$, measurement vectors $\mathbf{y}_i$ and  sparse vectors $\mathbf{x}_i$ are collected in the  training data. In this experiment, we consider matrices  with   elements   are drawn from zero mean, unit variance Gaussian distribution. The dimensions of the measurement matrix is selected as $M=40$ and $N=100$. The failure rate and relative mean square error of PC-SBL  is compared with L-SBL in Figures \ref{Block_Arbit_FR_1}  and \ref{Block_Arbit_MSE_1}. 
 The Figures  \ref{Block_Arbit_FR_1} and \ref{Block_Arbit_MSE_1}  show that the RMSE and failure rate of the L-SBL trained for arbitrary measurement matrix is slightly higher than the L-SBL network for a specific (randomly selected) measurement matrix.  Nonetheless, the L-SBL network trained for arbitrary measurement matrix outperforms PC-SBL algorithm. This indicates that when L-SBL is trained with a single measurement matrix $\mathbf{A}$, even though  the $\mathbf{A}$ is not used in the DNN,  the weights of the DNN do adapt to the structure of $\mathbf{A}$, yielding better recovery performance.  
\subsection{L-SBL with MMV Model}\label{Simulation_MMV_Model} 
\begin{figure}[t]
  	\centering	
  	\includegraphics[width=9cm,height=6cm]{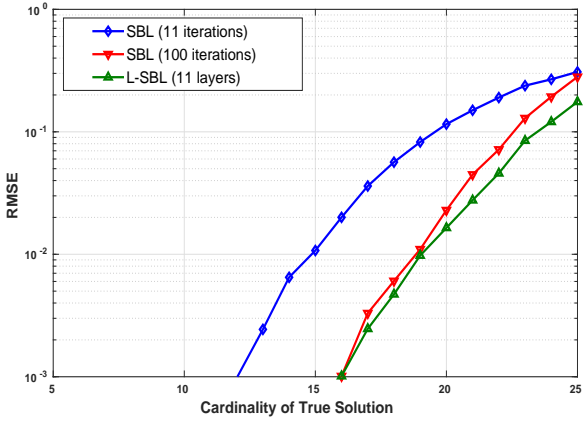}   	
  	\caption{Comparison of RMSE (Sparse Signal,  $L=3$)}
  	\label{Sparse_MMV_MSE_1}
  \end{figure}

\begin{figure}[t]
  	\centering	
  	\includegraphics[width=9cm,height=6cm]{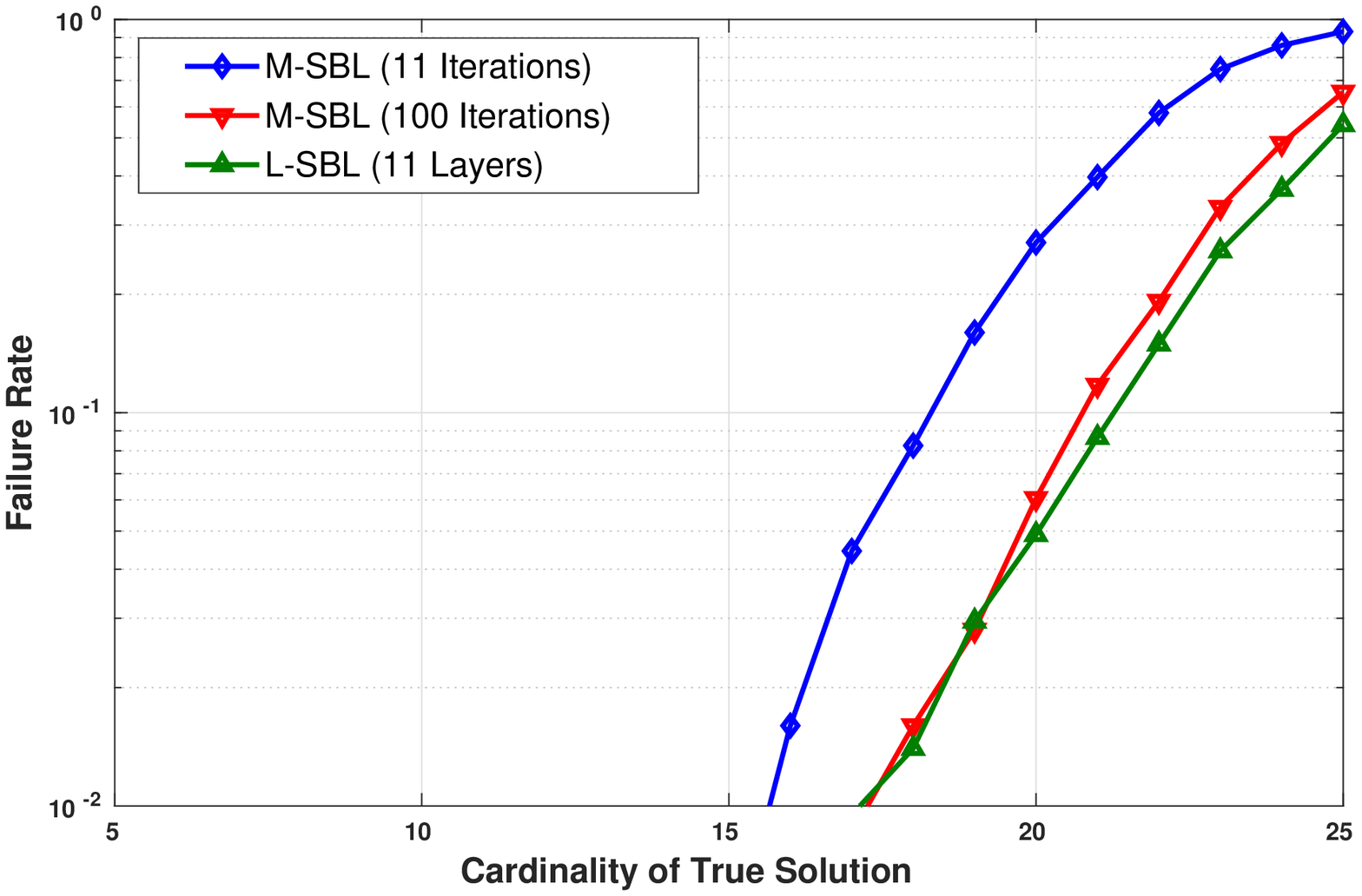} 	
  	\caption{Comparison of  Failure Rate (Sparse Signal,  $L=3$) }
  	\label{Sparse_MMV_FR_1}
  \end{figure}

In this subsection, we illustrate the potential of  L-SBL  to exploit multiple measurements for reducing mean square error and failure rate. First we compare the M-SBL algorithm with L-SBL for sparse signal recovery using the MMV model. The source vectors $\{\mathbf{x}_i\}_{i=1}^L$ are jointly sparse and the nonzero elements are independent and identically distributed.   The dimension of the measurement matrix  is chosen as $M=30$ and $N = 50$. The number of  measurements  $L$ is selected as  $3$. The M-SBL algorithm is compared with L-SBL  in Figures \ref{Sparse_MMV_MSE_1} and \ref{Sparse_MMV_FR_1}.   The plots show that L-SBL network  with 11-layers outperforms the M-SBL algorithm with $11$ iterations. The 11-layer L-SBL network shows comparable performance with  $100$ iterations of the SBL algorithm.

\begin{figure}[t]
	\centering    	
    \includegraphics[width=9cm,height=6cm]{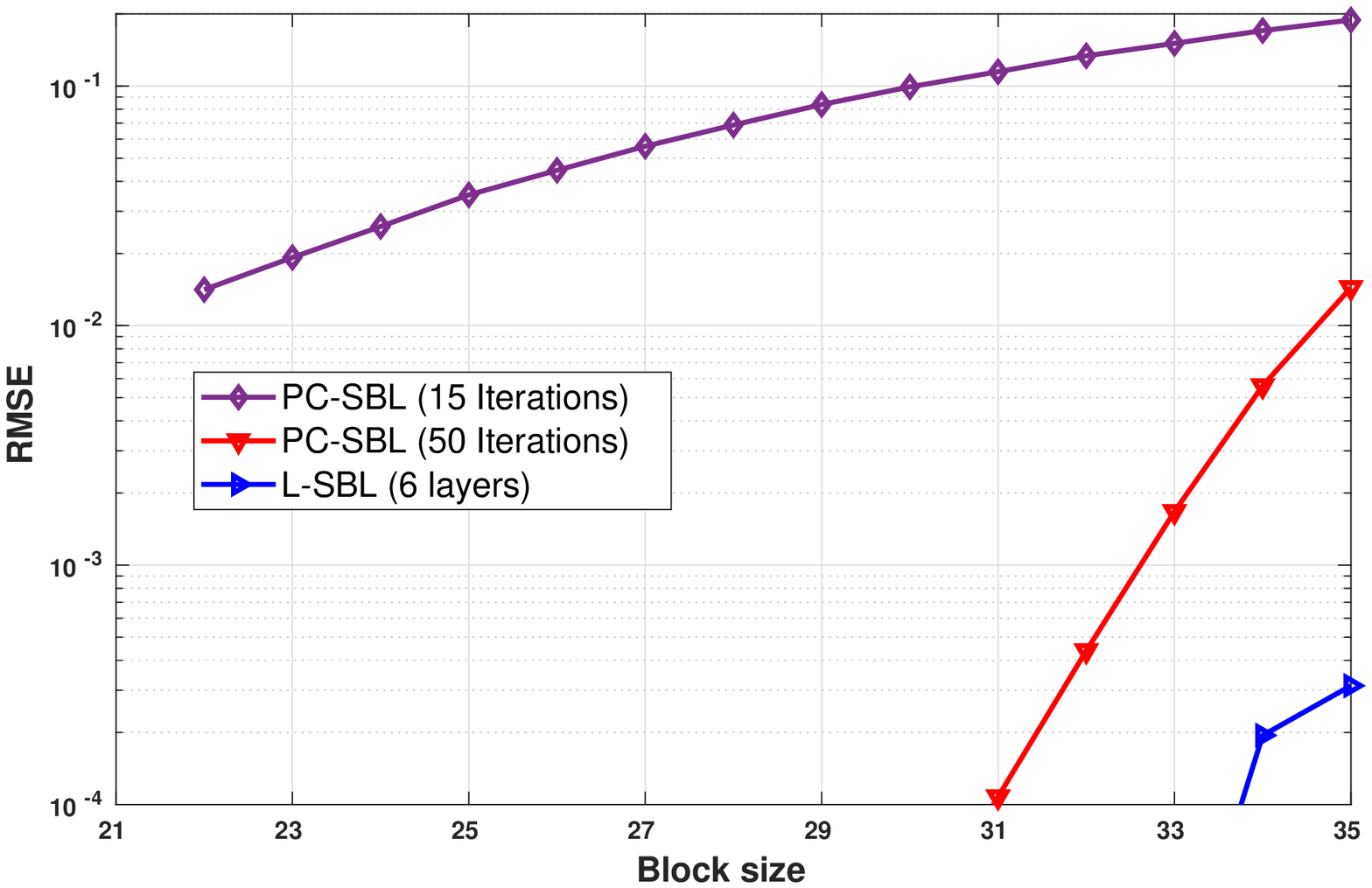} 
    \caption{Comparison of RMSE (Block-Sparse Signal, $L=3$)}
    	\label{BlockSparse_MMV_MSE_1}
    \end{figure}

\begin{figure}[t]
  	\centering  	
    	\includegraphics[width=9cm,height=6cm]{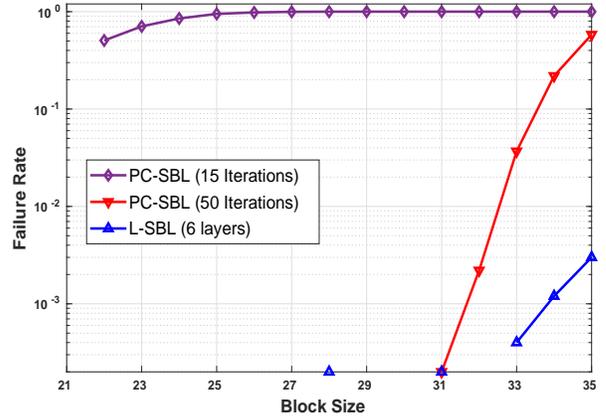} 	
   	\caption{Comparison of  Failure Rate (Block-Sparse Signal,$L=3$) }
    	\label{BlcokSparse_MMV_FR_1}
    \end{figure}

In Figures \ref{BlockSparse_MMV_MSE_1} and  \ref{BlcokSparse_MMV_FR_1},  we  evaluate the performance of  L-SBL to recover block sparse vectors using multiple measurements. In this comparison, we modified the original PC-SBL to utilize multiple measurement vectors. The computational complexity  of the PC-SBL with $15$ iterations is more than the L-SBL with six layers. The results show that once again, L-SBL with six layers  has reduced mean square error as well as failure rate than  PC-SBL with $50$ iterations.   
\subsection{L-SBL for source vectors with arbitrary pattern}\label{Simulation_Arbitrary_Pattern} 
\begin{figure}[t]
  	  	\centering	
  	\includegraphics[width=9cm,height=6cm]{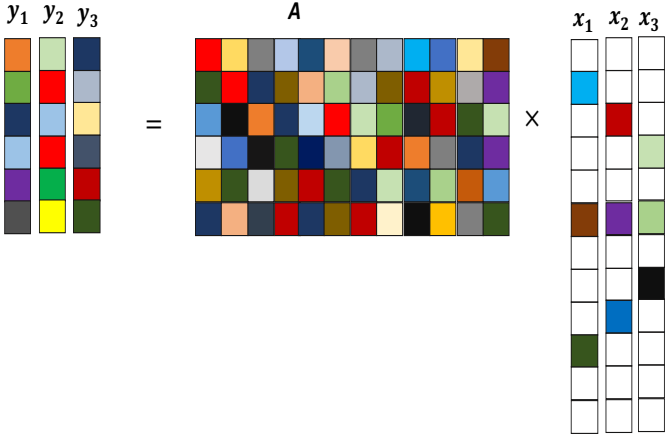}	
  	 	\caption{MMV model with arbitrary source pattern}
  	\label{MMV_Arbit_Model}
  \end{figure}

\begin{figure}[t]
       	\centering	     	
       	\includegraphics[width=9cm,height=6cm]{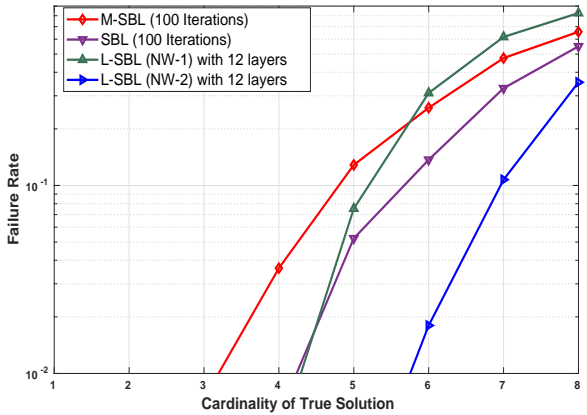} 	
       	\caption{Failure Rate (source vectors with arbitrary patterns) }
       	\label{Sparse_MMV_Arbit_FR_1}
       \end{figure}

\begin{figure}[t]
   	\centering	    	
     	\includegraphics[width=9cm,height=6cm]{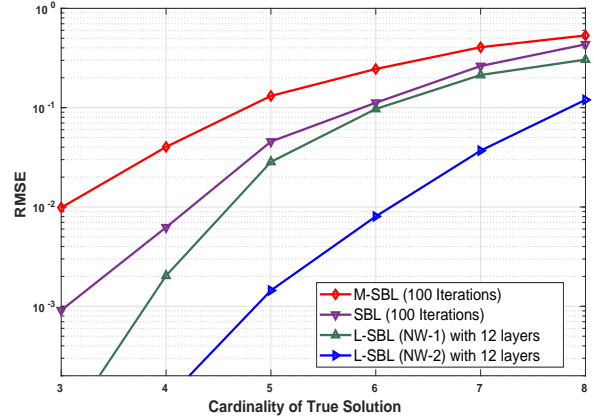} 	
       	\caption{ RMSE (source vectors with arbitrary patterns)}
     	\label{Sparse_MMV_Arbit_MSE_1}
     \end{figure}

Here we explore the performance of the L-SBL network, when the source vectors in multiple measurements are not jointly sparse. Consider the  patterns of nonzero elements among source vectors shown in Figure \ref{MMV_Arbit_Model}. These types of source patterns may arise, for example, the direction of arrival (DoA) estimation from multiple measurements.  Such patterns can exist in the scenarios where fast moving as well as stationary targets exist together in the radar's field of view. During the training of L-SBL network, training data set is generated according to the pattern shown in Figure \ref{MMV_Arbit_Model}. The trained L-SBL model learns an inverse function to recover sparse vectors  using the  patterns existing in the source vectors.   We compare the two architectures presented in section \ref{L-SBL Architecture} with SBL and M-SBL algorithms. The second architecture \emph{L-SBL (NW2)} has higher degrees of freedom due to the increased dimensions of the modified measurement matrix $\hat{\mathbf{A}}$ and signal covariance matrix $\mathbf{R}_x$. We consider a measurement matrix $\mathbf{A}$ with dimensions $M=15$ and $N=30$ in our simulation. The number of measurements $L$ is chosen as $3$.  The failure rate and relative mean square error of SBL and M-SBL algorithms are compared with L-SBL in Figure \ref{Sparse_MMV_Arbit_FR_1} and \ref{Sparse_MMV_Arbit_MSE_1}. 

The \emph{L-SBL (NW2)} shows superior  performance over other algorithms and \emph{L-SBL (NW1)}. Since the source vectors are not jointly sparse, performance of the M-SBL  algorithm is poor and the single measurement based SBL outperforms  M-SBL.  
\subsection{Weight matrices learned by L-SBL}\label{Weight_Matrix_Analysis}
\begin{figure}[t]
    	\centering	     	
       	\includegraphics[width=9cm,height=6cm]{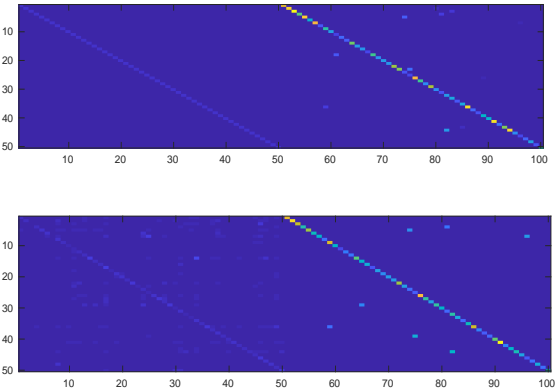} 	
       	\caption{L-SBL weight matrices for sparse  recovery (SMV)}
       	\label{Weight_L_SBL_SMV_Sparse}
       \end{figure}

\begin{figure}[t]
	  	\centering	         	
         	\includegraphics[width=9cm,height=6cm]{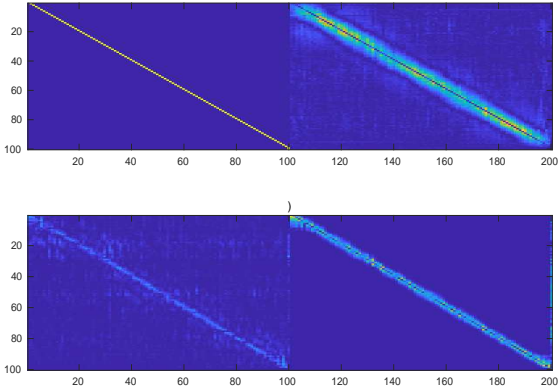} 	
        	\caption{L-SBL weight matrices for block-sparse  recovery}
         	\label{Weight_L_SBL_SMV_BlockSparse}
         \end{figure}

\begin{figure}[t]
	  	\centering	         	
         	\includegraphics[width=9cm,height=6cm]{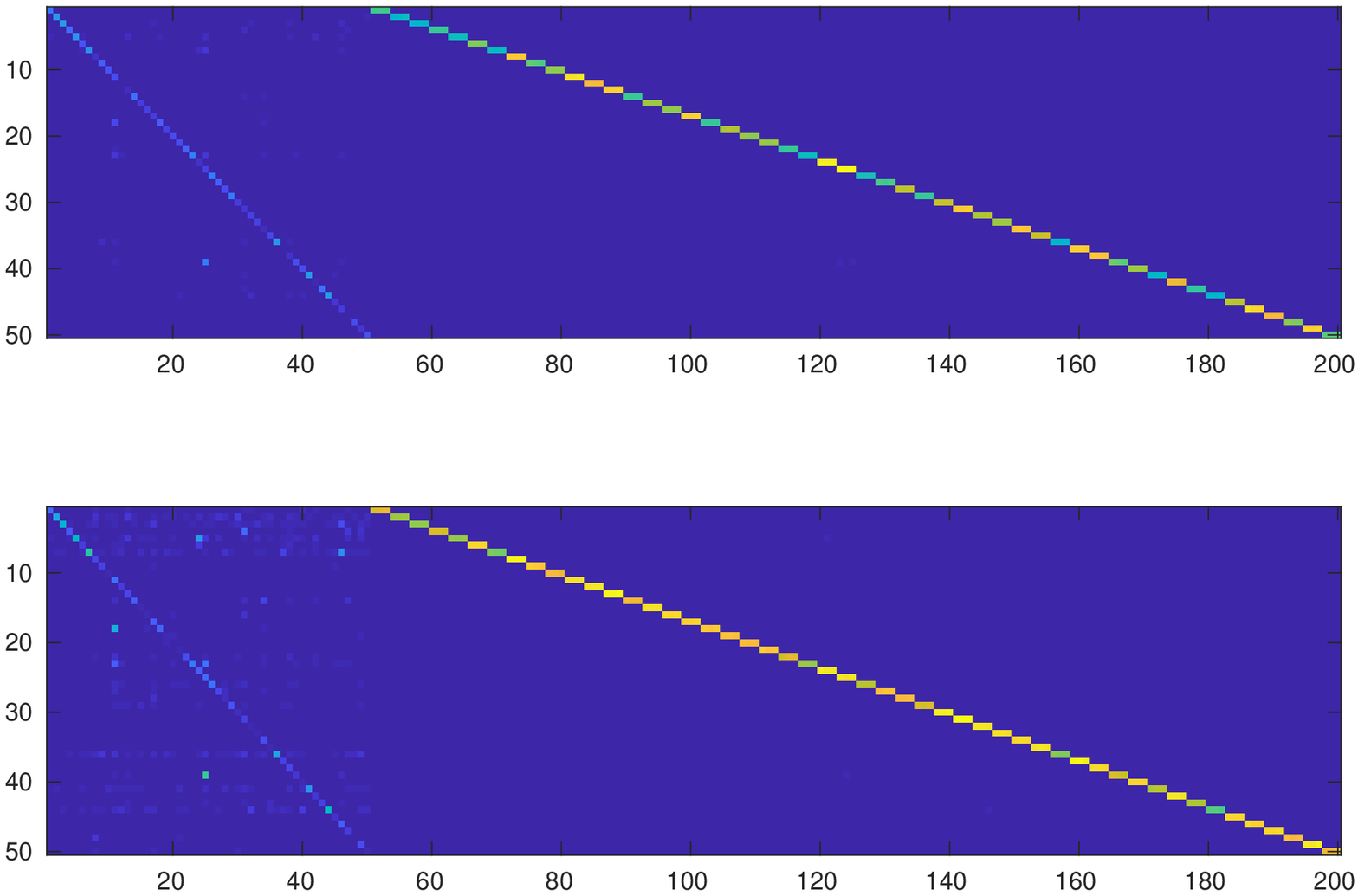} 	
         	\caption{L-SBL weight matrices for sparse recovery (MMV)}
         	\label{Weight_L_SBL_MMV_BlockSparse}
         \end{figure}

 We now present the weight matrices learned by the DNN in the training phase, which yields interesting insights into its superior performance.  In each iteration of the SBL algorithm, the estimate of hyperparameters $\{\alpha_i^t\}_{i=1}^N$ from the previous iteration outputs $\{(x_i^{t-1})^2\}_{i=1}^N$ and $\{[\mathbf{\Phi}^{t-1}]_{i,i}\}_{i=1}^N$can be implemented using a matrix multiplication given by
\begin{equation}
\\  \begin{bmatrix}
\frac{1}{	\alpha_{1}}\\
\vdots & \\
\frac{1}{\alpha_{N}}
\end{bmatrix}\\ =   \begin{bmatrix}
\mathbb{I}_{N}& \mathbb{I}_{N} \\
\end{bmatrix}\\
\\  \begin{bmatrix}
[\mathbf{\Phi}^{t-1}]_{1,1}\\
[\mathbf{\Phi}^{t-1}]_{2,2}\\
\vdots & \\
(x_1^{t-1})^2\\
\vdots & \\
(x_N^{t-1})^2\\
\end{bmatrix}\\
\label{eq:SBL_Implement}
\end{equation}
In numerical simulations, we used a single layer dense network for the estimation of hyperparameters $\{\alpha_i\}_{i=1}^N$. A single  layer  dense network  learns a weight matrix $\mathbf{W}$ and a bias vector $\mathbf{b}$  from the training data. The weight matrices  learned by the L-SBL  network are not the same as \eqref{eq:SBL_Implement}. The weight matrices implemented in  two different L-SBL layers for the recovery of a sparse vector from a single measurement vector are shown in Figure \ref{Weight_L_SBL_SMV_Sparse}, which indicates that different functions are implemented  in different  layers of the L-SBL network. Recall that the  nonzero elements of the  sparse vectors are drawn from uniform distribution which is different from the hierarchical Gaussian prior assumed by SBL. Such deviation from the assumed signal model may lead to the  improved performance of L-SBL network over SBL.    The weight matrices learned for the recovery of  block sparse vectors are also different from the weight matrices in single sparse vector recovery problem. The weight matrices of two  L-SBL layers in the block sparse signal recovery problem are shown in Figure \ref{Weight_L_SBL_SMV_BlockSparse}.  The learned weight matrices introduce a coupling between the adjacent elements of the sparse vector $\mathbf{x}$.  Moreover, the functions implemented in different layers of L-SBL are not the same.   The  weight matrices of L-SBL during the testing with MMV model are shown in Figure \ref{Weight_L_SBL_MMV_BlockSparse}, which illustrate that L-SBL exploits the joint sparsity among multiple source vectors in the estimation of hyperparameters. Finally, the weight matrices implemented during the training of samples with arbitrary patterns among source vectors are shown in Figure \ref{Weight_L_SBL_MMV_BlockSparse}. In this case, the off-diagonal elements of the second half of the weight matrix  are also nonzero. This indicates that L-SBL utilizes the patterns of the nonzero elements among source vectors to improve the recovery performance. Note that, in   \eqref{eq:SBL_Implement}, to estimate the hyperparameters $\{\alpha_i\}_{1=1}^N$,  the diagonal elements of the error covariance matrix and the sparse vector estimate from previous iteration are combined using equal weights. However, in the  learned weight matrix,  L-SBL network gives more importance to the sparse vector estimate from the previous layer for the estimation of hyperparameters.
 
         \begin{figure}[t]
	  	\centering	         	
         	\includegraphics[width=9cm,height=6cm]{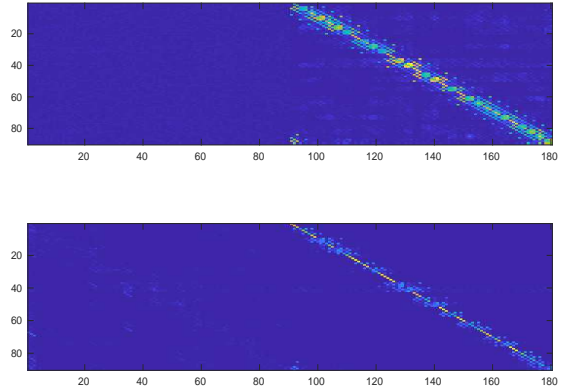} \caption{L-SBL weight matrices for sparse recovery (Arbitrary Pattern)}
         	\label{Weight_L_SBL_Arbitrary_BlockSparse}
         \end{figure}          
     
In this section,  we demonstrated the superior performance of L-SBL over other algorithms in the recovery of sparse vectors from single or multiple measurements.   The L-SBL network can recover block sparse vectors without the knowledge of block boundaries. Moreover, the weight matrices learned by the L-SBL exploits the arbitrary patterns of the nonzero elements among  source vectors to  reduce mean square error and failure rate. In the next section, we consider the localization of an extended target using a MIMO radar. The reflected  signal from an extended target can be modeled as  a block sparse vector and the introduced L-SBL network can be used in the recovery of the target signature.
 
\section{Extended target Detection using L-SBL}\label{Extended_Detection}
The detection of an extended target using radar or sonar can be modeled as a block sparse signal recovery problem. In \cite{shilpa_SPAWC_2016}, the  variational garrote approach  is extended for the recovery of a block sparse vector, where different block sizes are considered in different range bins. Since the proposed block sparse signal recovery scheme in \cite{shilpa_SPAWC_2016} assumes prior knowledge about block partitions, the algorithm is sensitive to the boundary mismatches. We demonstrate that the L-SBL can  detect the extended target from multiple radar sweeps, without knowledge of the block sizes or boundaries. 
\subsection{Signal model}
Consider a narrow band MIMO radar system   with $M_t$ transmitting antennas, $M_r$ receiving antennas and $L$ radar sweeps. Let $\mathbf{s}_i \in \mathbb{C}^{Q\times1} $ be the waveform emitted by the $i^{\text{th}}$ transmitting antenna. The set of doppler shifted waveforms from the $M_t$ antennas for the $d^{\text{th}}$ doppler bin  collected in the matrix $\mathbf{S}_d  \in \mathbb{C}^{M_t \times L} $ is,
\begin{equation}
\mathbf{S}_d = [\mathbf{s}_1(\omega_d), \mathbf{s}_2(\omega_d)......\mathbf{s}_{M_t}(\omega_d)]^T
\end{equation}
where  $\mathbf{s}_i(\omega_d) = \mathbf{s}_i \odot \boldsymbol{\psi}(\omega_d)$ is the doppler shifted waveform of the $i^{\text{th}}$ transmitting antenna and  $\psi(\omega_d) $ is given by 

\begin{equation}
\boldsymbol{\psi}(\omega_d)  = [1,e^{j\omega_d},......e^{j(Q-1)\omega_d}]^T
\end{equation}
The received sensor signal in $l^{\text{th}}$ radar sweep is given by

\begin{equation}\label{eqExt1}
\mathbf{Y}^{(l)} = \sum_{d=1}^{N_d}\sum_{r=1}^{N_r}\sum_{a=1}^{N_a}x_{d,r,a}^{l}\mathbf{b}(\theta_a)\mathbf{a}^T(\theta_a)\tilde{\mathbf{S}}_d\mathbf{J}_r + \mathbf{W}^{(l)} 
\end{equation}
where $N_a$ denotes the number of angular bins, $N_r$ denotes  the number of range bins and $N_d$ denotes the number of doppler bins. $\mathbf{b}(\theta_a) \in \mathbb{C}^{M_r \times 1} $ represents the steering vector of the receiver array towards the look angle $\theta_a$,  $\mathbf{a}(\theta_a) \in \mathbb{C}^{M_t \times 1} $ denotes the steering vector of the transmitter  array towards the look angle $\theta_a$ and $x_{d,r,a}^{l}$ is the scattering coefficient of the extended target. Further, $\tilde{\mathbf{S}}_d$ denotes the zero appended waveform matrix, which is given by
\begin{equation}
\tilde{\mathbf{S}}_d  = [\mathbf{S}_d, \mathbf{0}_{M_t \times N_r-1}] \in \mathbb{R}^{M_t \times(Q+N_r-1)},
\end{equation}
and $\mathbf{J}_r \in \mathbb{R}^{(Q+N_r-1) \times (Q+N_r-1)}$ is the shifting matrix that accounts for  the  different propagation times  of the returns from adjacent bins at the receiving array.
\begin{equation}
\mathbf{J}_r =
\begin{bmatrix}
\overbrace{0\ldots 01}^{r} & & \mathbf{0}\\ & \ddots & \\ \mathbf{0} & & 1 
\end{bmatrix}
\end{equation}	
The received signal in $l^{\text{th}}$ radar sweep given by \eqref{eqExt1}  can be rewritten as
\begin{equation}\label{ExtTarg1}
\mathbf{y}_c^{(l)} = \mathbf{A}_c\mathbf{x}_c^{(l)} + \mathbf{w}_c^{(l)}, 
\end{equation}
where $\mathbf{y}_c^{(l)} = vec(\mathbf{Y}^{(l)}) \in \mathbb{C}^{M_c \times 1}$, $M_c = M_r(Q+N_r-1)$, $N_c =N_aN_dN_r $ and 
$\mathbf{x}^l = [x_{1,1,2}^{l},...x_{N_d,N_r,N_a}^{l}] \in \mathbb{C}^{N_c \times 1} $.
The new measurement matrix $\mathbf{A}$ is  given by
\begin{equation}\label{ExtTarg2}
\mathbf{A}_c = [\mathbf{u}_{1,1,1},\mathbf{u}_{1,1,2}  ....\mathbf{u}_{N_d,N_r,Na}], 
\end{equation}
where $\mathbf{u}_{d,r,a} = vec(\mathbf{b}(\theta_a)\mathbf{a}^T(\theta_a)\tilde{\mathbf{S}}_d\mathbf{J}_r)$ and $\mathbf{w}^{(l)} = vec(\mathbf{W}^{(l)} )$. 
 We can arrange the  measurements of different radar sweeps $\{\mathbf{y}_c^{(l)}\}_{l=1}^L$  as columns of the matrix $\mathbf{Y}_c$.  Similarly, let $\mathbf{X}_c$ be the matrix with block sparse vectors $\{\mathbf{x}_c^{(l)}\}_{l=1}^L$ as its columns. Then, the signal model can be expressed as 

\begin{equation}\label{eq:ExtTarg3}
\mathbf{Y}_c = \mathbf{A}_c\mathbf{X}_c +\mathbf{N}_c
\end{equation}
where $\mathbf{N}_c$ represents the matrix with noise vectors  $\{\mathbf{w}_c^{(l)}\}_{l=1}^L$.
Now, many software packages for deep neural network implementation do not support the complex data type.  So we consider  an equivalent model in real vector space for extended target detection. We can express \eqref{eq:ExtTarg3} as follows:
\begin{equation}
\begin{split}
\mathbf{Y} &= \mathbf{A}\mathbf{X} + \mathbf{N},
\label{ExtendeTarget_Model}
\end{split}
\end{equation}
where the block sparse matrix $\mathbf{X}  \in \mathbb{R}^{N \times L}$ is $[Re(\mathbf{X}_c)^T,Im(\mathbf{X}_c)^T]^T $. The noise matrix $\mathbf{N}$ contains independent and identically distributed Gaussian random variables with zero mean and  variance $\sigma^2$. The measurement vector $\mathbf{Y}$ in the real vector space is given by
$\mathbf{Y} = [Re(\mathbf{Y}_c)^T,Im(\mathbf{Y}_c)^T]^T  \in \mathbb{R} ^{M \times L}$, 
and the measurement matrix $\mathbf{A}$ is related to $\mathbf{A}_c$ as
\begin{equation*}
\begin{split}
\\ \mathbf{A} =
\begin{bmatrix}
Re(\mathbf{A}_c) & -Im(\mathbf{A}_c) \\
Im(\mathbf{A}_c) & Re(\mathbf{A}_c)
\end{bmatrix}\\
\end{split}
\end{equation*}
where $Re()$ and $Im()$ denote the real part and imaginary part of the matrix. The dimensions of  measurement matrix in real vector space $\mathbf{A}$  is related to the dimensions of $\mathbf{A}_c$  as  $M = 2M_c$  and $N=2N_c$.

\subsection{Numerical Results}
\begin{figure}[t]
	\includegraphics[width=9cm,height=4.5cm]{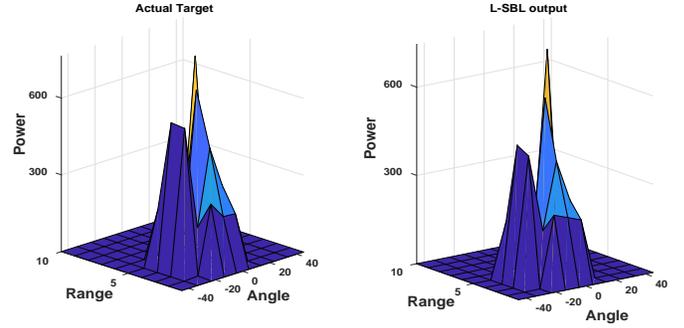} 	
	
	\caption{Target Detected by L-SBL}
	\label{ExtendedTarget_Realization}
\end{figure}

\begin{figure}[t]
	\includegraphics[width=9cm,height=6cm]{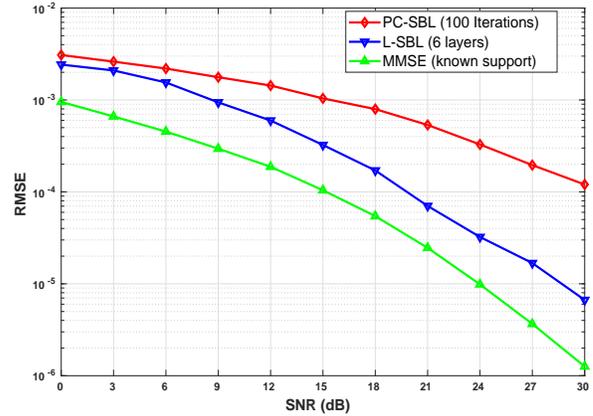} 	
	
	\caption{RMSE vs SNR (Extended Target Detection)}
	\label{ExtendedTarget_RMSE_SNR}
\end{figure}

In numerical simulations, we consider a MIMO radar system with number of transmitter antennas $M_t =2$. Number of receiver antennas $M_r=10$. Ten angular bins are considered between $-45^{\degree}$ to $45 ^{\degree}$. The antenna  spacing in the receiver as well as  transmitter array is chosen as   $\lambda/2$ , where $\lambda$ denotes the wavelength corresponding to the  signal transmitted by the radar. The number of doppler bins $N_d$ is chosen as one and the transmitted waveform is selected as the Hadamard code  of length $Q =2$.  The number of radar sweeps $L$ is chosen as two, the scattering coefficients of the extended target is drawn from the standard complex Gaussian distribution.  We use a synthetically generated training data set according to  \eqref{ExtendeTarget_Model} to train the L-SBL network. The training procedure of the L-SBL network is as described in Algorithm \ref{Alg1}. The SNR of the received block sparse vector is also chosen randomly between $0$ dB and $ 30$ dB in each training sample. The expression of the SNR in the testing as well as training data is,        
\begin{equation}
\textbf{SNR} =  \frac{\mathbb{E}(\mathrm{Tr}{(\mathbf{X}^T\mathbf{A}^T\mathbf{A}\mathbf{X}))}}{\mathbb{E}(\mathrm{Tr}{(\mathbf{N}^T\mathbf{N}))}}
\end{equation} 
A realization of the target detected by L-SBL in 30 dB SNR  is shown in Figure \ref{ExtendedTarget_Realization}.
We compare the performance of L-SBL  with PC-SBL. We also compare  with the Minimum Mean Square Estimator (MMSE) with known support, which has the least mean square error.
In Figure \ref{ExtendedTarget_RMSE_SNR}, we compare the relative mean square error (RMSE) of different algorithms against SNR. The plot shows that the relative mean square error of L-SBL with $8$ layers is less than the PC-SBL algorithm with $100$ iterations.

\section{Conclusion}\label{Conclusion}
In this paper, we presented a new deep learning architecture named as Learned-SBL for sparse signal processing. The L-SBL network can recover sparse or block sparse vectors depends on the training of the network. The L-SBL utilizes multiple measurements  to reduce failure rate and mean square error. The arbitrary patterns of the non-zero elements among multiple source vectors are also exploited by L-SBL to enhance the performance.   The introduced learned L-SBL avoids the retraining  of the network in the applications where measurement matrix changes with time. We compared L-SBL with other algorithms and showed the application in the detection of an extended target using a MIMO radar.
\ifCLASSOPTIONcaptionsoff
  \newpage
\fi



%
\bibliographystyle{IEEEtran}
\bibliography{IEEEabrv,bib_report}



%

%
%





\end{document}